\documentclass[sigconf]{acmart}
\usepackage{booktabs} 
\usepackage{multirow}
\usepackage{tabularx}
\usepackage{enumitem}
\usepackage{arydshln}
\usepackage{subcaption}
\usepackage{amsfonts}
\usepackage{amsmath}
\usepackage{amssymb} 
\usepackage{algorithm}   
\usepackage[noend]{algcompatible}
\usepackage{mathtools}
\usepackage{bm}
\usepackage{microtype}
\usepackage[american]{babel}

\def\b{{\bf b}}
\def\c{{\bf c}}

\def\e{{\bf e}}

\def\v{{\bf v}}

\def\E{{\bf E}}

\def\P{{\bf P}}
\def\Q{{\bf Q}}

\def\W{{\bf W}}

\def\0{{\bf 0}}
\def\1{{\bf 1}}
\def\2{{\bf 2}}
\def\3{{\bf 3}}
\def\4{{\bf 4}}
\def\5{{\bf 5}}
\def\6{{\bf 6}}
\def\7{{\bf 7}}
\def\8{{\bf 8}}
\def\9{{\bf 9}}

\DeclareMathOperator*{\argmin}{argmin}
\newcolumntype{R}{>{\raggedleft\arraybackslash}X}
\newcolumntype{C}{>{\centering\arraybackslash}X}
\AtBeginDocument{%
  \providecommand\BibTeX{{%
    \normalfont B\kern-0.5em{\scshape i\kern-0.25em b}\kern-0.8em\TeX}}}

\begin{document}

\title{Learning Multi-granular Quantized Embeddings for Large-Vocab Categorical Features in Recommender Systems}

\author{Wang-Cheng Kang, Derek Zhiyuan Cheng, Ting Chen, Xinyang Yi, Dong Lin, Lichan Hong, Ed H. Chi}
\email{wckang@ucsd.edu, {zcheng, iamtingchen, xinyang, dongl, lichan, edchi}@google.com}
\affiliation{%
  \institution{Google, Inc}
  \institution{UC San Diego}
}
\renewcommand{\shortauthors}{W.-C. Kang et al.}

\begin{abstract}
  Recommender system models often represent various sparse features like users, items, and categorical features via embeddings. A standard approach is to map each unique feature value to an embedding vector. The size of the produced embedding table grows linearly with the size of the vocabulary. Therefore, a large vocabulary inevitably leads to a gigantic embedding table, creating two severe problems: (i) making model serving intractable in resource-constrained environments; (ii) causing overfitting problems. In this paper, we seek to learn highly compact embeddings for large-vocab sparse features in recommender systems (recsys). First, we show that the novel Differentiable Product Quantization (DPQ) approach can generalize to recsys problems. In addition, to better handle the power-law data distribution commonly seen in recsys, we propose a Multi-Granular Quantized Embeddings (MGQE) technique which learns more compact embeddings for infrequent items.
  To the best of our knowledge, we are the first to tackle end-to-end learning of quantized embeddings for recommender systems. Extensive experiments on three recommendation tasks and two datasets show that we can achieve on par or better performance, with only ~20\% of the original model size.
\end{abstract}

\ccsdesc[500]{Information systems~Recommender systems}
\ccsdesc[500]{Information systems~Collaborative filtering}
\ccsdesc[300]{Information systems~Data compression}
\keywords{Recommender Systems, Model Compression, Quantized Embeddings}



\maketitle

\begin{figure}[t]
    \centering
    \includegraphics[width=\linewidth]{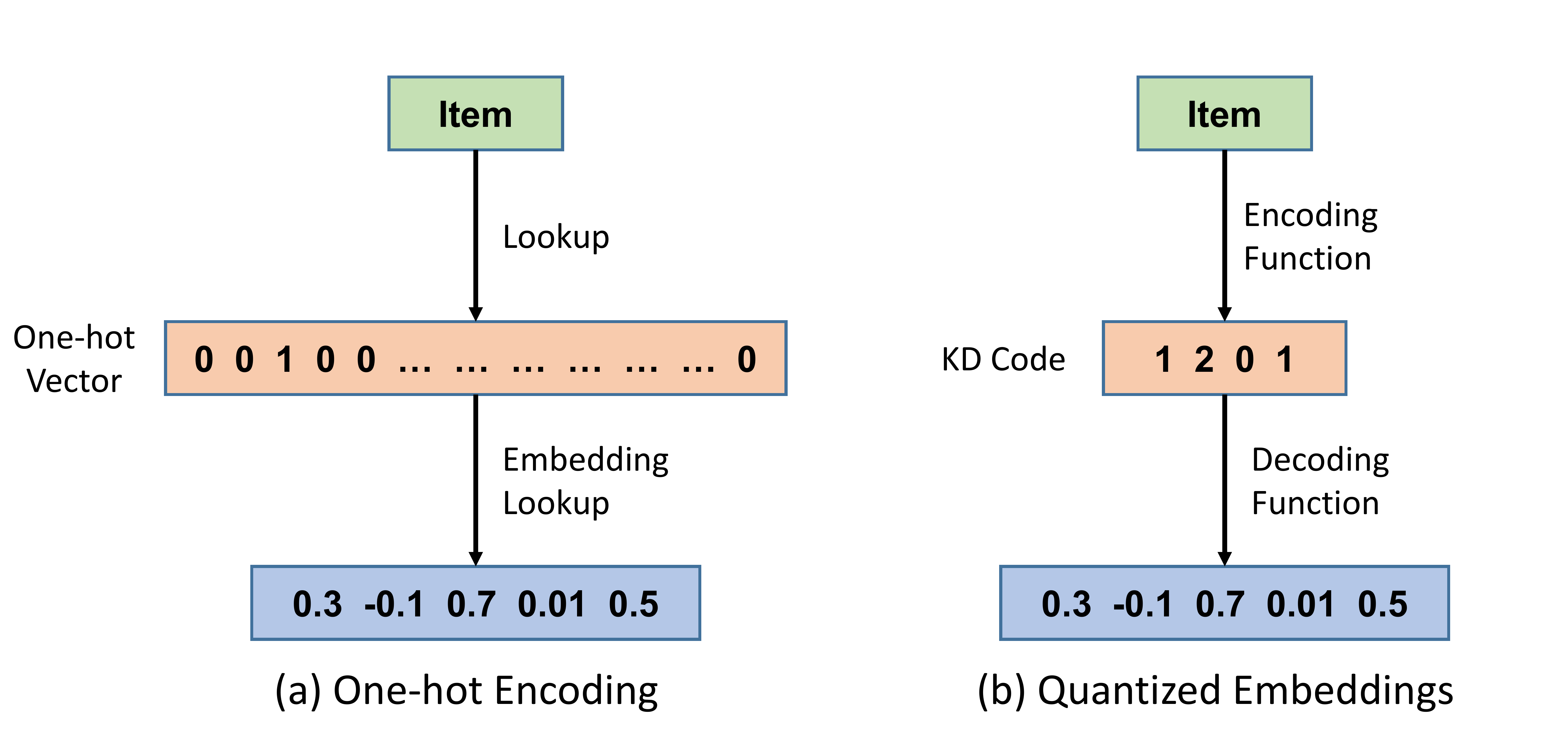}
    \caption{An illustration of the embedding lookup procedure in one-hot encoding and quantized embeddings~\cite{DBLP:conf/icml/ChenMS18, DBLP:journals/corr/abs-1908-09756}. Quantized embeddings encode items with learned discrete codes, and apply a decoding function to get the real-valued embeddings. The encoding and decoding functions are learned end-to-end with the target task.}
    \label{fig:intro}
\end{figure}

\section{Introduction}
Representation learning for categorical features has been a very active research area over the last two decades~\cite{rumelhart1986learning, DBLP:conf/nips/BengioDV00,DBLP:conf/nips/MikolovSCCD13}. The quality of the learned representations is crucial to the overall model quality. In linear models, sparse features are often represented as one-hot vectors. This approach works well to help models memorize, however fails to generalize and introduces lots of parameters by learning coefficients for all coordinates of the one-hot vector.

In deep neural network models, categorical features are often represented as embeddings, and these embeddings are essential to both memorization and generalization. Embeddings are heavily used in fully-connected nets~\cite{DBLP:conf/www/HeLZNHC17, Zhao:2019:RVW:3298689.3346997}, recurrent neural nets~\cite{DBLP:journals/corr/BahdanauCB14, Wu:2017:RRN:3018661.3018689}, and transformer models \cite{transformer, DBLP:conf/naacl/DevlinCLT19}. An embedding function maps items\footnote{For convenience, in this paper, we use the term `items' to represent elements (e.g., users, items, and categorical features) in the vocabulary.} into a continuous space (i.e., the one-hot encoding), as shown in Figure~\ref{fig:intro}. The learned embeddings for different items are directly comparable since they are mapped to the same high dimensional latent space. This significantly helps improve the generalization capability for deep neural nets.

Although the one-hot encoding is very powerful, learning efficient and effective embeddings for categorical features is  challenging, especially when the vocabulary for the sparse features is large, and the training data is highly skewed towards popular items. The size of the embedding table grows linearly with the vocabulary size, leading to two severe problems: (i) making model serving intractable in resource-constrained environments; (ii) causing overfitting problem due to the over-parameterization. This is still the case even for industrial recommender systems with fairly sufficient computing power. For example, in a recent work about YouTube Recommendation~\cite{DBLP:conf/recsys/YiYHCHKZWC19}, it's unveiled that tens of millions of parameters are used to learn embeddings for YouTube video IDs alone. This requires very sophisticated distributed training and serving strategies, and makes the model vulnerable to overfitting. Moreover, the distribution for large-vocab sparse features is often power-law skewed, which significantly hurts the embedding quality learned for torso and long-tail items as there are significantly fewer training examples than those of the head items.

To cut down model size from embeddings, there have been some works on embedding compression, such as the hashing tricks \cite{DBLP:conf/icml/WeinbergerDLSA09}, low-rank factorization~\cite{DBLP:journals/corr/abs-1909-11942}, and quantization~\cite{DBLP:journals/corr/abs-1908-09756}. On the other hand, to learn better torso and tail item embeddings, there is a line of works seeking to allocate more embedding capacity to frequent words and reduces the capacity for less frequent words with the benefit of reducing overfitting to rare words~\cite{DBLP:conf/nips/ChenSLCH18, DBLP:conf/iclr/BaevskiA19}. Inspired by the two lines of work, we seek to propose a compact embedding scheme with variable capacities. Moreover, many of these efforts were focused on natural language understanding tasks. The embedding compression problem for recommendation tasks remains to be fully studied.

 In this paper, we show that the quantization-based embedding compression method Differentiable Product Quantization~(DPQ)~\cite{DBLP:journals/corr/abs-1908-09756} can generalize to recsys tasks. Moreover, we propose Multi-granular Quantized Embeddings~(MGQE) which extend DPQ with variable embedding capacities to adapt to highly skewed data distributions in recsys tasks. MGQE significantly reduces the model size with on par or better performance compared to the full model. 
 
 Our main contributions are as follows:

\begin{itemize}[leftmargin=5mm]
    \item We propose multi-granular quantized embeddings~(MGQE), a novel embedding compression method  designed for power-law distributed data. MGQE adopts highly compact quantized representations to reduce the model size, and allocates less capacities to long-tail items to further cut down the storage space. To the best of our knowledge, MGQE is the first quantized embedding based approach with variable capacities.
    \item We apply MGQE to the embedding compression problem in large-scale recomemnder systems, which is critical but not well studied in the existing literature. In recommendation settings, we examine both classic compression techniques and recent methods originally designed for compressing NLP models.
    \item We perform extensive experiments to compress several representative recommendation models for three recommendation tasks on both academic and industrial datasets. Our results show that the proposed method MGQE outperforms existing approaches, and can generally reduce 80\% of the original model size while matching or improving the full model performance.
\end{itemize}

\section{Related Work}

We first introduce model compression techniques mainly designed for compressing hidden layers in neural networks, followed by embedding compression techniques that are closely related to our work. Lastly, we briefly introduce compact recommendation models based on binary user/item representations.

\subsection{Model Compression} Though deep learning approaches have gained tremendous success in computer vision~\cite{DBLP:conf/nips/KrizhevskySH12,DBLP:conf/cvpr/HeZRS16} and natural language processing~\cite{DBLP:journals/corr/WuSCLNMKCGMKSJL16, DBLP:conf/naacl/DevlinCLT19}, one significant problem is the huge model size mainly due to the parameters (e.g., weights) in hidden layers. The gigantic model size makes the training and deploying of deep models much harder in memory or computation constrained environments like GPUs or mobile devices. Hence, model compression becomes an active research area with various kinds of techniques being adopted. For example, Deep Compression~\cite{DBLP:journals/corr/HanMD15} prunes small-weight connections and adopts quantized representations and Huffman encoding to further reduce the storage space. XNOR-Net~\cite{DBLP:conf/eccv/RastegariORF16} adopts binary convolutional filters for acceleration and compression. Knowledge distillation~\cite{DBLP:journals/corr/HintonVD15,DBLP:conf/kdd/TangW18} seeks to learn a smaller student model from the supervision of a trained larger teacher model. Low-rank factorization~\cite{DBLP:conf/icassp/SainathKSAR13} and the hashing trick~\cite{DBLP:conf/icml/ChenWTWC15} have also been adopted for compressing hidden layers.

However, existing model compression methods generally focus on compressing hidden layers in deep neural networks like AlexNet~\cite{DBLP:conf/nips/KrizhevskySH12} and ResNet~\cite{DBLP:conf/cvpr/HeZRS16}, which significantly differ from recommendation models. Existing neural recommendation models usually contain a shallow network (e.g., less than 4 hidden layers)~\cite{DBLP:conf/recsys/Cheng0HSCAACCIA16, DBLP:conf/www/HeLZNHC17,DBLP:conf/kdd/YingHCEHL18}, which only accounts for a small portion of the model size. Hence, directly applying these methods to compress hidden layers in recommendation models will not significantly reduce the model size.

\subsection{Embedding Compression}
Learning to embed sparse features (e.g., words, items, etc.) into a high dimensional space has become the de facto approach in domains like natural language processing and recommender systems ~\cite{DBLP:conf/nips/MikolovSCCD13, DBLP:conf/recsys/BarkanK16, DBLP:conf/kdd/GrbovicC18}.
However, the vocabulary size could be quite large in recommender systems, and the embedding table becomes gigantic and accounts for most of the parameters in the model. Hence, it is necessary to adopt more compact embedding schemes to cut down the model size. However, the embedding compression problem has not been extensively explored, especially for recommender systems. A line of work uses random hash functions to map the vocabulary to a smaller number of hashed buckets, and hence the storage space is significantly reduced~\cite{DBLP:conf/icml/WeinbergerDLSA09, DBLP:conf/recsys/SerraK17, DBLP:conf/nips/SvenstrupHW17}. Such random-hashing based methods fail to capture the inherent similarities between different items, and the hash codes may not be amenable to the target task, which results in potential compromise on model performance.

Recently, a few researchers used learned hash codes to compress word embeddings for NLP tasks, and showed better performance than random codes~\cite{DBLP:conf/icml/ChenMS18,DBLP:conf/iclr/ShuN18}. However, the hash codes in these methods are pretrained or learned with distillation which uses pre-trained embeddings as a guideline. To this end, Differentiable Product Quantization~(DPQ)~\cite{DBLP:journals/corr/abs-1908-09756} proposes a differentiable quantization framework that enables end-to-end training for embedding compression, and achieves significant compression ratio on NLP models. Other than representing each word with compact embeddings, another line of work explores how to efficiently allocate variable embedding capacities to frequent and infrequent words~\cite{DBLP:conf/iclr/BaevskiA19, DBLP:conf/nips/ChenSLCH18}. We seek to exert the advantages of the two lines of work via proposing MGQE, an end-to-end embedding compression approach with quantized embeddings and variable capacities.

\begin{table*}
\centering

\begin{tabular}{lcccc}
\toprule
  &\multicolumn{2}{c}{\textbf{Training}} &\textbf{Serving}\\
          & \textbf{Embedding Size} &  \textbf{End-to-end?}& \textbf{Embedding Size} \\ \midrule
\textbf{One-hot Encoding}                                         &  $32nd$   & Yes       &$32nd$ \\
\textbf{Low-Rank Factorization}                          &  $32nr+32rd$    &   Yes    & $32nr+32rd$   \\
\textbf{Scalar Quantization}                             &  $32nd$      &  No     &    $ndb$     \\   
\textbf{Differentiable Product Quantization~(DPQ)~\cite{DBLP:journals/corr/abs-1908-09756}}       &  $32nd+32Kd$     &   Yes    &$nDlogK+32Kd$  \\
\textbf{Multi-Granular Quantized Embeddings~(MGQE)}      &  $32nd+32Kd$     &   Yes     & $\sum_{i=1}^m |V_i|DlogK_i+32Kd$     \\       
\bottomrule
\end{tabular}
\caption{Embedding size comparison for different compression approaches. The model size is the number of bits needed to store the model. We assume that single-precision floating-point (i.e., float32) is used, which costs 32 bits to store a real value.}\label{tab:model}
\vspace{-0.3cm}\end{table*}

\begin{table}[t]

\begin{tabularx}{\linewidth}{lX}
\toprule
Notation&Description\\
\midrule
$n\in \mathbb{N}$  & vocabulary size\\
$d\in \mathbb{N}$  & embedding dimension\\
$D\in \mathbb{N}$  & number of subspaces\\
$K\in \mathbb{N}$  & number of centroids\\
$r\in \mathbb{N}$  & the rank in low-rank factorization\\
$b\in \mathbb{N}$  & the number of bits in scalar quantization\\

$\widetilde V=(V_1, V_2,..., V_m)$  & multi-tier partition in MGQE, where $m$ is the number of groups \\
$\widetilde D=[D_1,D_2,...,D_m]$  & variable numbers of subspaces for each group, where $D_i$ is for the $i$-th group\\
$\widetilde K=[K_1,K_2,...,K_m]$  & variable numbers of centroids for each group, where $K_i$ is for the $i$-th group\\
\bottomrule
\end{tabularx}
\caption{Notation. \label{tb:notation}}

\vspace{-0.3cm}\end{table}

\subsection{Compact Recommendation Models} To achieve efficient retrieval for recommendation, an orthogonal line of work represents user/item vectors in a Hamming space~\cite{DBLP:conf/cvpr/LiuHDL14, DBLP:conf/sigir/ZhangSLHLC16,DBLP:journals/corr/abs-1909-05475,DBLP:conf/kdd/LianLG00C17}, and thus generating item recommendations can be efficiently done in constant or sub-linear time~\cite{DBLP:journals/pami/0002PF14}. While using binary vectors can also reduce the storage space, the accuracy of such models is generally worse compared to real-valued models for two reasons: (i) such models are highly constrained; and (ii) they may lack sufficient flexibility when aiming to precisely rank the Top-K items. In contrast, although quantized embeddings store the discrete codes for each item, they will be dequantized and operated in a flexible continuous space. Moreover, these binary recommendation models are not generic model compression approaches that can be used for compressing real-valued recommendation models.

\section{Embedding Learning for Sparse Features}

Embedding learning has become a standard technique to handle sparse features like words, entities, and users/items. The core idea is to map discrete items into a learned continuous $d$-dimensional space. Then, these learned embeddings are either fed into simple scoring functions (e.g., inner product), or neural networks (e.g., RNNs) to generate the final predictions. Specifically, we have an embedding function $\mathcal{T}: V \to R^d$ mapping an item (from vocabulary $V$ with size $|V|=n$) to a $d$-dimensional embedding vector $\v$. Generally, the embedding function can be decomposed into two components: $\mathcal{T}=f\circ g$, where $g$ is an encoding function to represent discrete objects in a discrete space, and $f$ is a decoding function to generate the continuous embedding $\v$. In this section, we briefly introduce several options of the encoding and decoding functions ($f$ and $g$). The notation is summarized in Table~\ref{tb:notation}.

\subsection{One-hot Encoding} The one-hot encoding approach assigns each unique individual item with one dimension in a $n$-dimensional vector, where $n$ is the vocabulary size. Then each of these dimensions gets uniquely mapped into a $d$-dimensional embedding vector in the embedding table. This is equivalent to the following: (1) we apply the encoding function $g$ to encode item $s^{(i)}\in V$ with a one-hot encoding vector: $g(s^{(i)})=\b^{(i)} \in \{0, 1\}^{n}$ where $b^{(i)}_i = 1$ and $b^{(i)}_j = 0 (j\neq i)$; (2) we then apply the decoding function $f$, a learnable linear transformation $\W\in \mathbb{R}^{n\times d}$ to generate the embedding vector $\v$, that is, $\v=f(\b) = \W^T\b$.

Despite its simplicity, such an encoding scheme has several issues. First, the code utilization is highly inefficient, as it only utilizes $n$ codes while $n$-dimensional hamming space can represent $2^{n}$ different items. Specifically, the code utilization rate is almost zero with a large vocabulary: $n/2^{n}\to 0$ when $n\to \infty$. Second, the embedding table grows linearly with the vocabulary size and the embedding dimensionality: $n\times d\times 32$ bits with float32. For example, if we build a movie recommendation model with 1 billion movies with one-hot encoded embeddings with 100 dimensions, the embeddings will require 400GB (1,000,000,000 * 100 * 32 bits) of memory to store. The gigantic model size is not amenable for resource-constrained environments (e.g., on-device processing) and can lead to severe overfitting. Last but not least, it endows all items with the same embedding capacity, while real-world recommendation data typically follows power-law distributions and hence a large chunk of the items has few observation data.

\subsection{Low-rank Factorization}

Low-rank factorization is a classic matrix factorization method which assumes there is a low-rank latent structure in the matrix that could be compressed. In our case, we can factorize the embedding matrix $\W$ with $\W = \P\Q$ where $\P\in \mathbb{R}^{n\times r}$, $\Q\in \mathbb{R}^{r\times d}$, where $r$ is the rank (usually $r<d$). That is, $g$ is still the one-hot encoding function $g(s^{(i)})=\b^{(i)}$, and $f$ contains two linear transformations: $f(\b^{(i)})=\Q^T\P^T\b^{(i)}$. Hence, low-rank factorized embeddings could also be trained end-to-end with gradient descent.

The embedding size of such an approach is $O(nr+rd)$, which is generally smaller than the size of vanilla one-hot encoding $O(nd)$. Recently, low-rank factorization has been adopted in compressing embedding matrices in NLP models \cite{DBLP:journals/corr/abs-1909-11942}, and hidden layers in neural networks~\cite{DBLP:conf/icassp/SainathKSAR13}. However, we found low-rank factorization on the embeddings could hurt model performance in recsys models. Presumably because the embedding dimension (e.g., 10s to low 100s) in recommendation models is generally much smaller than that in NLP models (e.g.,, 728 for BERT~\cite{DBLP:conf/naacl/DevlinCLT19}, and 2000 for Word2Vec~\cite{DBLP:conf/nips/MikolovSCCD13}), and thus the rank cannot be further reduced.

\begin{figure}[t]
    \centering
    \includegraphics[width=\linewidth]{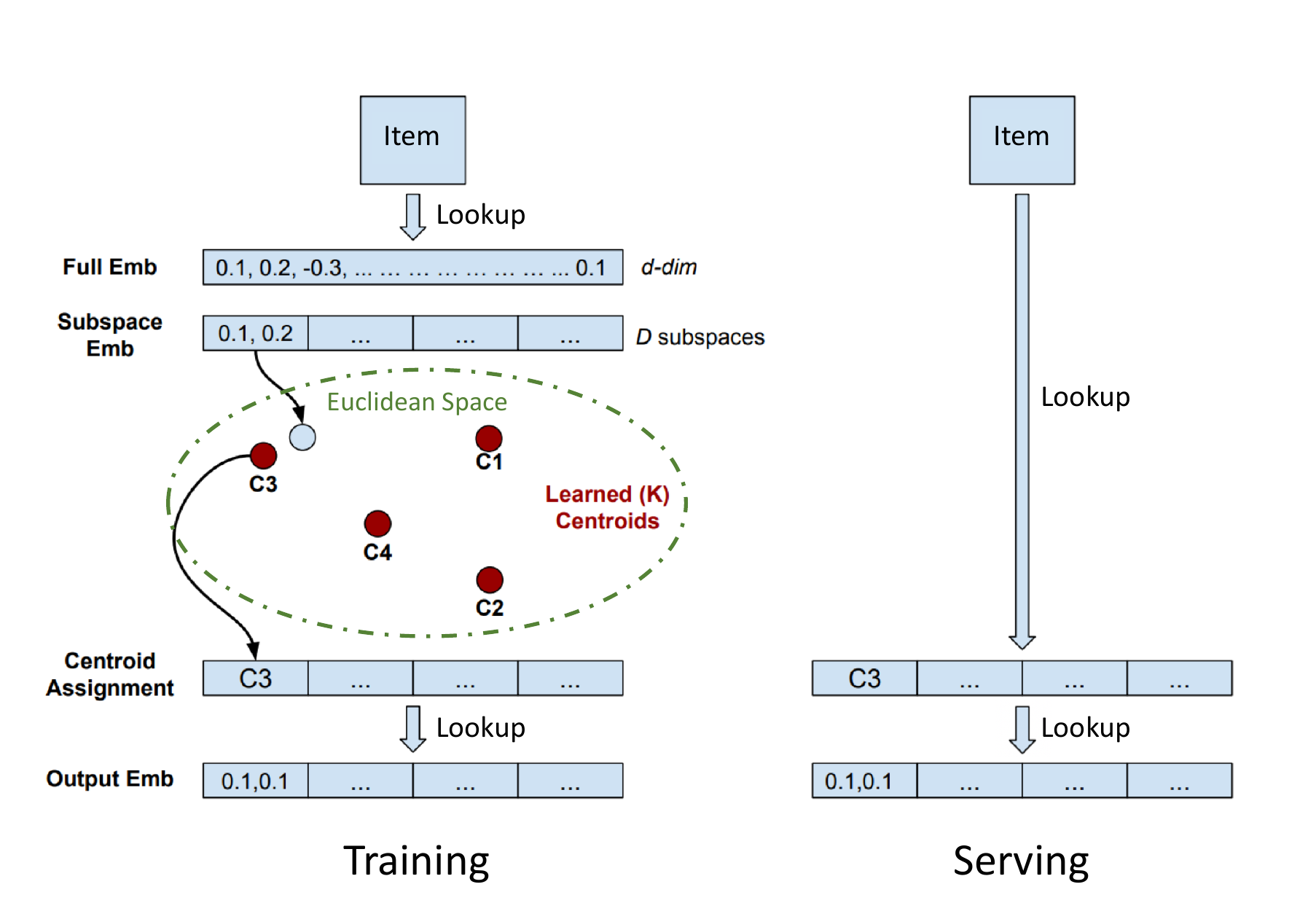}
    \caption{An illustration of the DPQ embedding lookup process during training and serving. The full embeddings are introduced to aid training, and are discarded during serving. Hence we only need to store the codes and the centroid embedding table during serving, which significantly reduces the model size.}
    \label{fig:dpq}
\end{figure}

\subsection{Differentiable Product Quantization}

Differentiable Product Quantization~(DPQ)~\cite{DBLP:journals/corr/abs-1908-09756} is a recent end-to-end embedding compression approach based on a highly compact encoding scheme called KD encoding \cite{DBLP:conf/icml/ChenMS18}. Unlike one-hot encoding which maps items to a partially utilized hamming space (i.e., $g:V\to \{0,1\}^{n}$), the KD encoding function is defined as $g: V\to \{1,\dots, K\}^D$.
For example, an item may be encoded as (1-2-3-1) when $K$=3 and $D$=4. Instead of adopting a pre-defined assignment (e.g., one-hot encoding) or random mapping (e.g., the hashing trick~\cite{DBLP:conf/icml/WeinbergerDLSA09}), the encoding function $g$ in DPQ is differentiable, and can be end-to-end trained with the target task, which allows adaptive assignment of similar codes to semantically similar items. In this paper, we focus on the vector quantization variant of DPQ~\cite{DBLP:journals/corr/abs-1908-09756}, as we found the softmax based variant does not work well in our preliminary study.

During training, DPQ aims to learn KD codes for items in the vocabulary. The model maintains an embedding vector $\e=[\e^{(1)};...;\e^{(D)}]\in\mathbb{R}^{d}$ for each item, where $D$ is the number of subspaces and $\e^{(i)}\in \mathbb{R}^{d/D}$ is the embedding vector in $i$-th subspace. In the $i$-th subspace, the model also maintains $K$ learnable centroid embeddings $\c_j^{(i)}\in \mathbb{R}^{d/D}$, where $j$=$1,...,K$. The KD codes are computed through a product quantization process, which contains two processes as follows. In the first encoding process, we find the index of the nearest centroid in each subspace:
\[g(\e)=(\argmin_k\|\e^{(1)}-\c^{(1)}_k\|, ..., \argmin_{k}\|\e^{(D)}-\c^{(D)}_k\|).\]
And then a decoding function simply retrieves centroid embeddings based on the codes, and concatenates them as the final embedding:
\[f(k_1,...,k_D)=[\c^{(1)}_{k_1}; ...; \c^{(D)}_{k_D}],\]
where $(k_1,...,k_D)=g(\e)$ are computed KD codes for the item. Although the overall encoding/decoding process $f\circ g$ is not differentiable due to the \emph{argmin} operation. DPQ addresses this issue and makes the process fully differentilable using the straight-through estimator~\cite{DBLP:journals/corr/BengioLC13}.


At serving time, the embedding vector $\e$ is discarded, as we only need to store the codes $(k_1,...,k_D)$ for each item and the centroid embeddings $\{\c_j^{(i)}\}$. We directly apply the decoding function $f$ on the codes to retrieve the final embedding. Hence, we only need $nD\log(K)$ bits to store code assignments for each item, and $K*D*d/D*32=32Kd$ bits to store the centroid embeddings. Typically, the code assignment is the dominant term as it grows linearly with the vocabulary size $n$. Figure~\ref{fig:dpq} depicts the lookup process in DPQ.

\section{Multi-granular Quantized Embeddings}

\begin{figure*}
\centering
\begin{subfigure}[b]{0.247\linewidth}
\includegraphics[width=\linewidth]{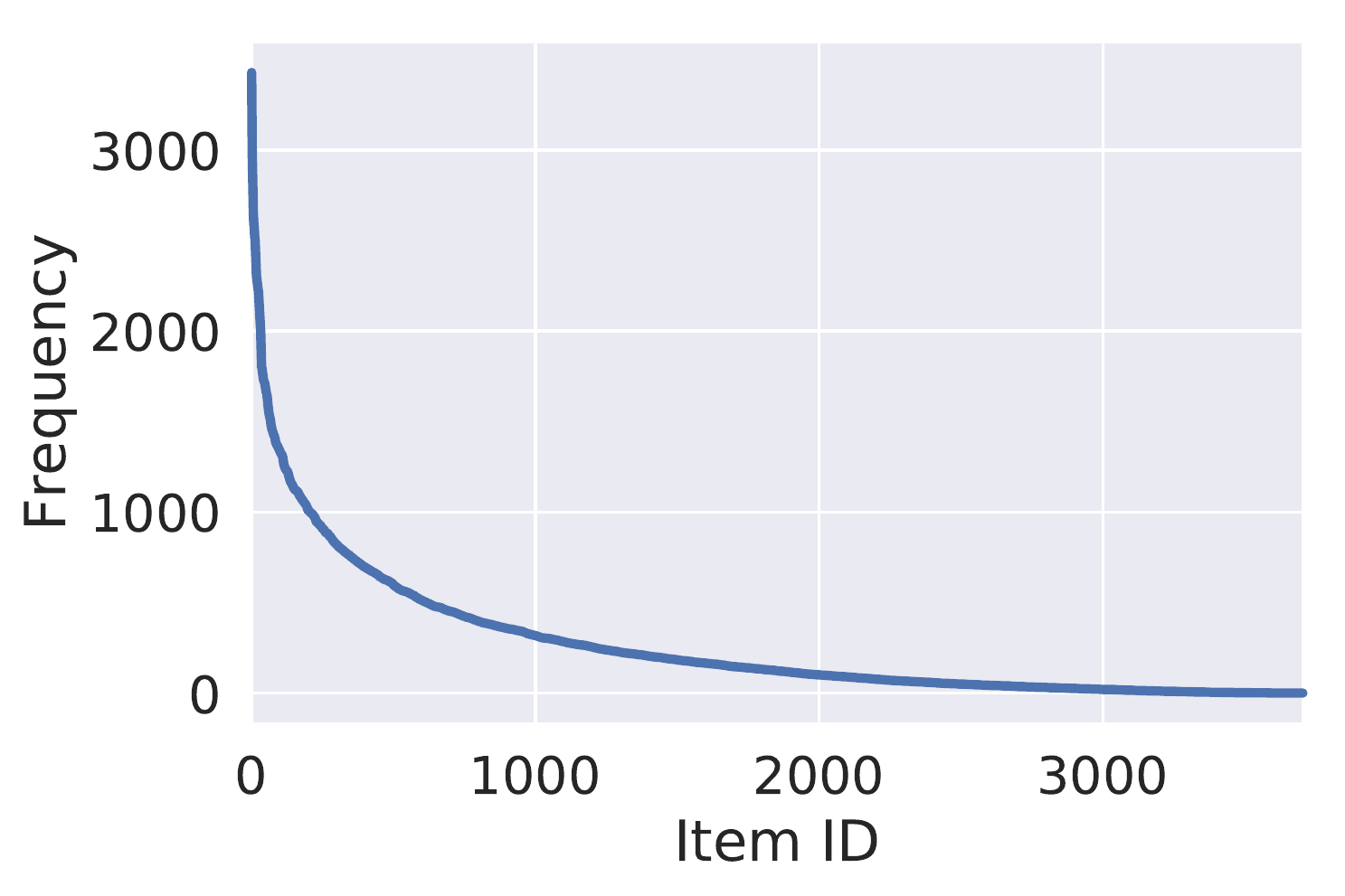}
\subcaption{MovieLens-1M}
\end{subfigure}
\begin{subfigure}[b]{0.247\linewidth}
\includegraphics[width=\linewidth]{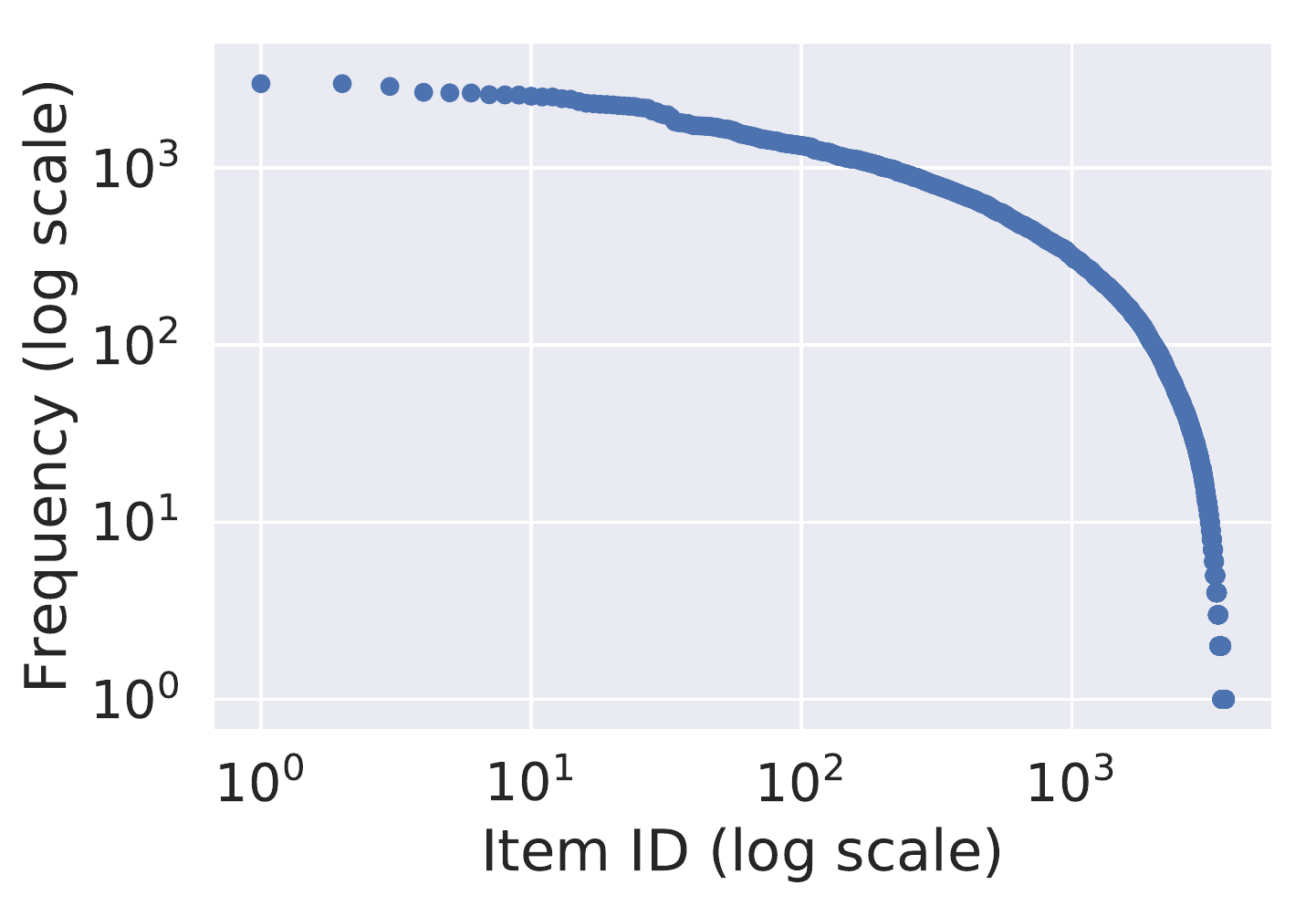}
\subcaption{MovieLens-1M (log scale)}
\end{subfigure}
\begin{subfigure}[b]{0.247\linewidth}
\includegraphics[width=\linewidth]{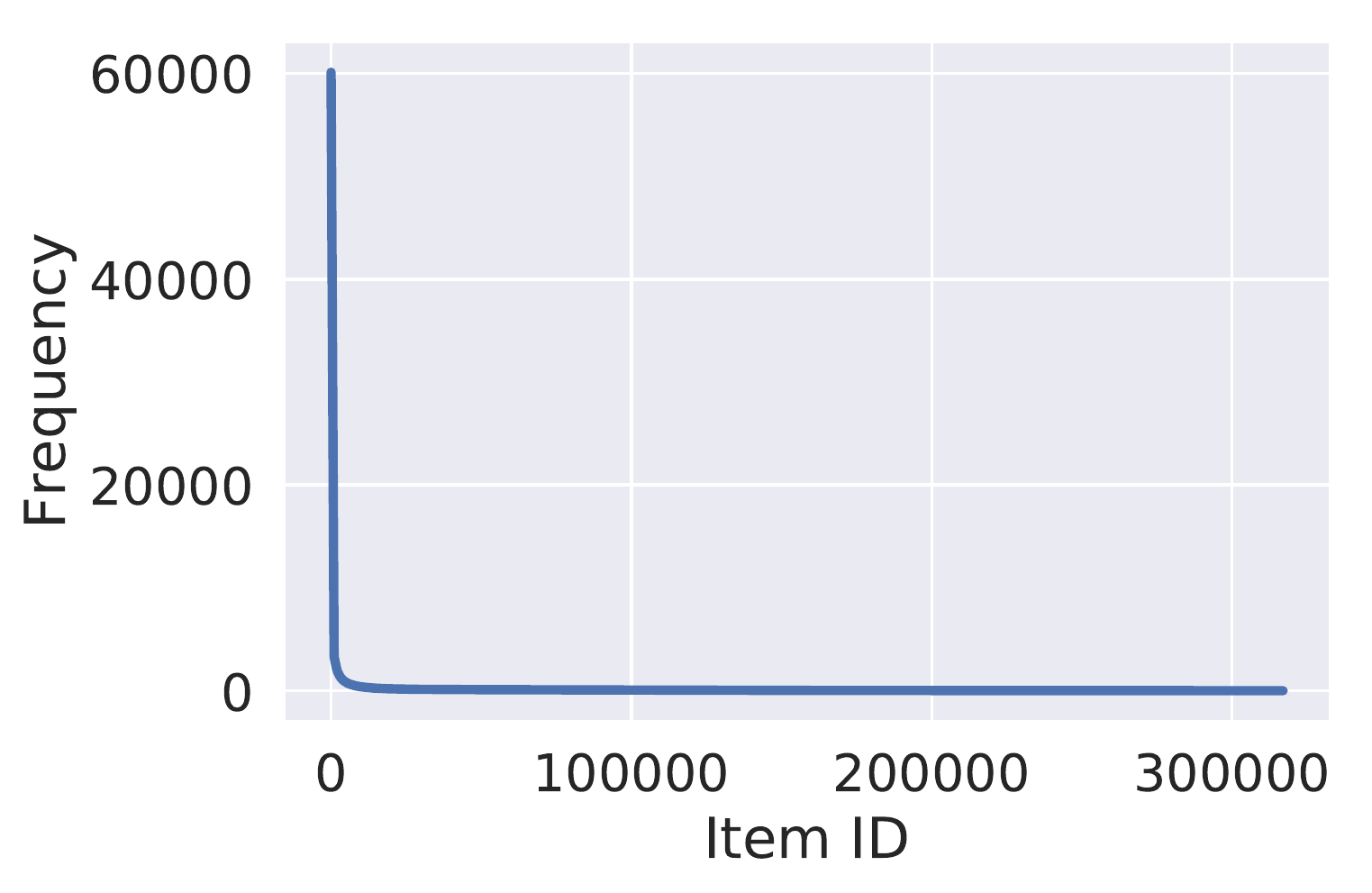}
\subcaption{AAR}
\end{subfigure}
\begin{subfigure}[b]{0.247\linewidth}
\includegraphics[width=\linewidth]{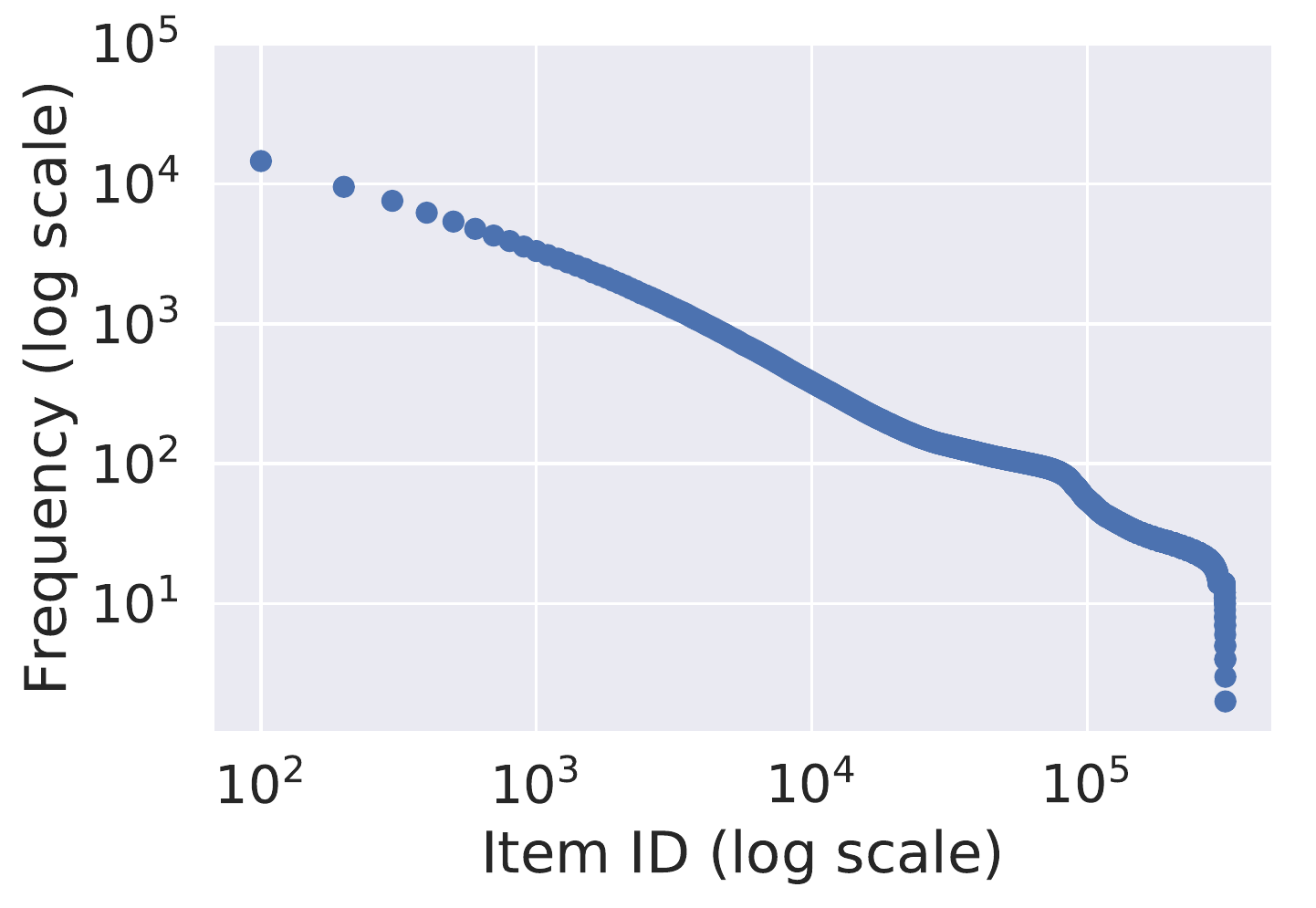}
\subcaption{AAR (log scale)}
\end{subfigure}
\caption{Dataset distributions.}\label{fig:data}
\end{figure*}

With a lower intrinsic dimensionality $D$ and quantized representations, DPQ significantly reduces the model size. However, by assigning same code capacity to each user/item, DPQ is unaware of the highly skewed power-law distributions as shown in typical recsys datasets (Figure~\ref{fig:data}). In recommendation domains, usually a few popular items dominate the training data, while the majority of the items (i.e. long-tail items) are rarely observed. In this case, allocating the same embedding capacity to all items is sub-optimal, as it could lead to overfitting on infrequent users/items due to data sparsity and high embedding dimensions. Moreover, we may not be able to learn fine-grained embeddings for tail items due to the limited data, and hence it is memory inefficient to store tail items' embeddings with the same memory space as head items. This motivates us to treat frequent and infrequent items differently via learning more compact embeddings for tail items. To this end, we propose Multi-granular Quantized Embeddings~(MGQE) which learns embeddings with different capacities for different items. MGQE adopts the quantized embedding DPQ as its underlying embedding scheme for two reasons: (1) DPQ is a highly compact end-to-end embedding learning approach, and achieves excellent compression performance; (2) we found that the quantized encoding scheme (i.e., KD encoding~\cite{DBLP:conf/icml/ChenMS18}) is highly flexible in terms of supporting different capacities (e.g., varying $D$ and $K$).

\subsection{Frequency-based Partitions}

The first question is how to split items into several groups to which we allocate different embedding capacities. We adopt an intuitive approach that partitions the items based on frequency (i.e., how many times an item appears in the training set). The intuition is that popular items frequently appear in the training set and have more associated observations, and hence we may need a large capacity for them to learn fine-grained embeddings. The frequency-based partition has been recently adopted for NLP models~\cite{DBLP:conf/nips/ChenSLCH18, DBLP:conf/iclr/BaevskiA19}, though our work differs in domains (i.e., recommendation models) and underlying embedding approaches (quantized embeddings).

We first assign ascending IDs to the items from the most popular to least popular ones. Then we split the frequency ordered vocabulary $V$ into a multi-tier partition $\widetilde V=(V_1,V_2, ..., V_m)$, where $\bigcup_{i=1}^{m}V_i=V$, $V_i\bigcap V_j=\emptyset$ for $i\neq j$, $m$ is the number of groups, $V_1$ contains the most popular items, and $V_m$ contains the least popular items. As we consider recommendation datasets which usually follow power-law distributions, typically we have $|V_1|<|V_2|<...<|V_m|$. The partition is heuristically set based on dataset statistics, as in~\cite{DBLP:conf/iclr/BaevskiA19, DBLP:conf/nips/ChenSLCH18}.

\subsection{Multi-granular Capacities with Quantized Embeddings}

The KD encoding scheme is highly flexible in terms of model sizes, as we can vary the embedding capacity by adjusting $D$ (the number of subspaces) or $K$ (the number of centroids). Hence we can directly derive two variants for the multi-granular capacity allocation via varying K or D. Specifically, instead of using a single $K$ and $D$, we extend DPQ via using a vector to represent the capacity allocation for each group: $\widetilde K=[K_1,...,K_m]$ and $\widetilde D=[D_1,...,D_m]$ where $K_1\geq K_2\geq...\geq K_m$ and $D_1\geq D_2\geq...\geq D_m$. Then we learn DPQ embeddings with $K_i$ and $D_i$ for items in $V_i$. For simplicity, we consider two variants: (1) \textbf{variable $\widetilde K$}: using fixed number of subspace $D$ with variable $\widetilde K$ for each group; (2) \textbf{variable $\widetilde D$}: using fixed number of centroids $K$ with variable $\widetilde D$ for each group. In this way, we allocate multi-granular embedding capacities (and storage space) for items with different popularity.

\textbf{A toy example:} Assuming that we have MGQE embeddings with $m=2$, $D=4$, and $\widetilde K=[16, 8]$, we will create a DPQ embedding table with $D=4$ and $K=16$ for items in $V_1$, and another DPQ embedding table with $D=4$ and $K=8$ for items in $V_2$. When retrieving the embedding for an item, we first check which group ($V_1$ or $V_2$) it belongs to, and then lookup in the corresponding DPQ embedding table.

However, a potential drawback of the two variants above is that we need to maintain a private centroid embedding table for each group, which increases the model size ($O(\sum_{i=1}^{m}K_id)$) for both training and serving. Hence, we propose a multi-granular scheme with centroids shared among groups. That is, we maintain a \emph{single} DPQ embedding table with $D$ and $K$. For items in the $i$-th group, it can only use \emph{first} $K_i$ centroids, instead of all $K$ centroids. In this way, we achieve the multi-granular embedding flexibilities via varying $K$, without additional storage cost. Moreover, we further reduce the model size compared to DPQ, as we only need $D\log(K_i) (i>1)$ bits to store the code assignment for a tail item (which accounts for majority of the items). We refer to this variant as \textbf{shared, variable $\widetilde K$}.

By default, we adopt the variant with shared centroids and varying numbers of centroids (\textbf{shared, variable $\widetilde K$}). In the experiments, we show the performance of all variants, and find that using shared centroids leads to satisfactory performance. Note that if we adopt the varying numbers of subspaces (i.e., $ D$), we cannot have a shared centroid approach, as the subspace dimensions would be different. A comparison of the embedding size of various approaches is summarized in Table~\ref{tab:model}.


\begin{figure}[t]
    \centering
    \includegraphics[width=\linewidth]{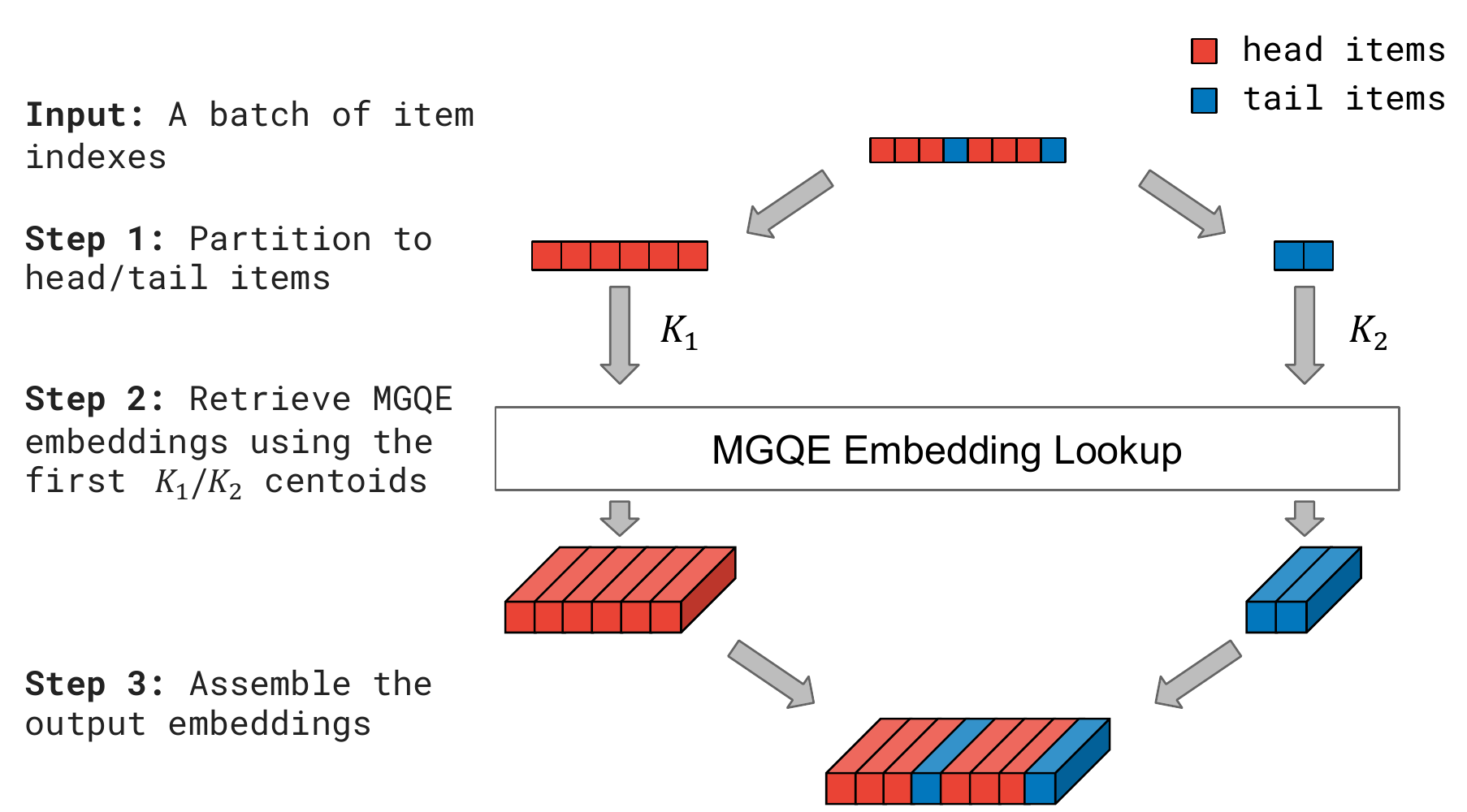}
    \caption{An example of the embedding lookup process of MGQE with two groups.}
    \label{fig:MGQE}
\end{figure}

\begin{algorithm}[t]
\caption{The group-wise embedding look-up process in MGQE.}\label{algo}
\label{algo:mih2}
\begin{algorithmic}
\STATE {\textbf{Hyper-parameters:} partition $\widetilde V=(V_1,V_2,...,V_m)$, embedding dimension $d$, number subspace $D$, variable numbers of centroids $\widetilde K$}
\STATE {\textbf{Initialization:} intialize a DPQ embedding class with $D$ and $K$}
\STATE {\textbf{Input:} a batch of items $S$=($s_1$, $s_2$,\dots,$s_B$)}

\STATE {Split $S$ into $m$ groups $G_1,G_2,...G_m$ according to the partition $\widetilde V$}
\FOR{$\mathit{i} = 1 \to \mathit{m}$}
  \STATE{$\E_i\gets$ MGQE\_embedding\_lookup\footnotemark($G_i$, $K_i$)}
\ENDFOR
\STATE {$\E\gets$ concatenate($\E_1,...,\E_m$) }
\STATE {Reorder $\E$ such that the $i$-th row of $\E$ is the embedding for item $s_i$}
\STATE {\textbf{return} $\E\in\mathbb{R}^{B\times d}$}
\end{algorithmic}
\end{algorithm}
\footnotetext{We extend the embedding look-up procedure in DPQ with an additional parameter $K_i$, which means we only use the first $K_i$ centroids when searching the nearest centroid.}

\subsection{Optimized Group-wise Embedding Lookup}

Unlike existing embedding learning methods (e.g., one-hot encoding, DPQ, etc.) that perform the same operation for items in a batch to retrieve embeddings, MGQE needs to apply different operations based on which groups they belong to. This results in a problem for MGQE: how to efficiently obtain embeddings for a batch of items? A straightforward approach is to process the items separately. That is, we apply a \emph{map} operation\footnote{e.g., tf.map\_fn in Tensorflow API.} with function $p$ on all the items in a batch, where $p$ identifies which group the given item belongs to, and retrieves the embedding with corresponding configurations (e.g., the corresponding DPQ embedding table). Hence, this will lead to $B$ (batch size) queries to retrieve DPQ embeddings. However, we found this is highly inefficient and makes the training process with MGQE much slower than that with DPQ embeddings. The main reason is that batch-wise matrix multiplication operation is highly optimized on GPU, but processing items separately (i.e., vector-matrix multiplication) can not leverage such acceleration.

However, as the embedding lookup process (e.g., centroid embedding table to be queried, use the first $K_i$ centroids for items in $V_i$, etc.) is the same for items in the same group, we propose to process the items by groups. Specifically, for a batch of items, we first split them into $m$ groups based on which groups they belong to. Then we obtain the embeddings for each group via $m$ queries of retrieving DPQ embeddings. As the number of groups $m$ is typically small, we found that this implementation of MGQE delivers similar training speed as the vanilla DPQ. Algorithm~\ref{algo} summarizes the lookup process of MGQE, and Figure~\ref{fig:MGQE} provides an example of embedding lookup with two groups.

\subsection{End-to-end Embedding Compression}

Similar to DPQ, MGQE can be trained end-to-end with the target task. This makes MGQE a very generic method as it can replace the embedding layers in models and can still be trained with gradient descent. The end-to-end learning can also help us learn better quantization guided by the target task, compared with conventional two-step quantization approaches like scalar quantization and product quantization~\cite{jegou2010product}.

In this paper, we focus on applying MGQE on recommendation models by simply replacing the user and item embedding layers with MGQE. As described in the following section, we evaluate MGQE on three representative recommendation models (GMF~\cite{DBLP:conf/www/HeLZNHC17}, NeuMF~\cite{DBLP:conf/www/HeLZNHC17}, SASRec~\cite{DBLP:conf/icdm/KangM18}) to show the generality of MGQE. Our method can also be applied to side categorical features (e.g., category, ad campaign id, etc.) which are also prevalent in recommender systems and CTR prediction~\cite{DBLP:conf/kdd/WangFFW17,DBLP:conf/kdd/LianZZCXS18}, though we plan to investigate this in the future.


\section{Experiments}

In this section, we present our experimental setup and results to investigate the following research questions:

\begin{description}
\item[\textbf{RQ1:}] Do the quantized embedding approaches (DPQ and MGQE) significantly reduce model size while matching or improving the performance with full embeddings?
\item[\textbf{RQ2:}] Does MGQE outperform DPQ and other embedding compression methods on power-law distributions in recsys tasks?
\item[\textbf{RQ3:}] What are the effects of the three multi-granular schemes for MGQE?
\item[\textbf{RQ4:}] Is the convergence of MGQE stable and as fast as that of full embeddings?
\item[\textbf{RQ5:}] Do learned codes uncover semantic information (i.e., do semantically similar items share similar codes)?

\end{description}

\begin{figure*}
    \centering
    \includegraphics[width=\linewidth]{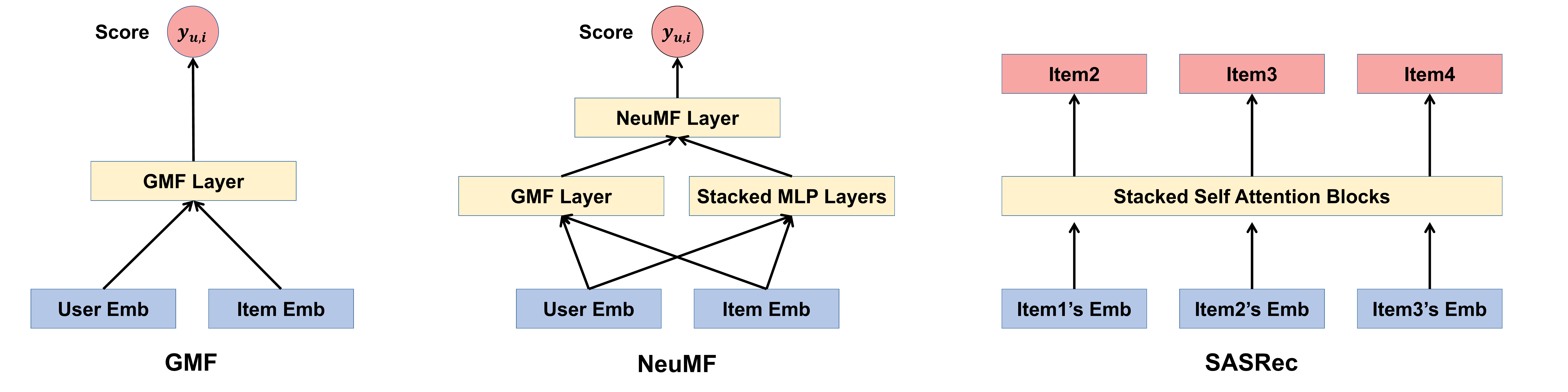}
    \caption{A simplified illustration of the three backbone recommendation models used in our experiments. GMF and NeuMF~\cite{DBLP:conf/www/HeLZNHC17} estimate user-item preference scores, and SASRec~\cite{DBLP:conf/icdm/KangM18} considers action sequences and seeks to predict the next item at each time step. We apply embedding compression on the input embeddings (blue rectangles in the figure) of the models. }
    \label{fig:models}
\end{figure*}

\subsection{Datasets}

We use two datasets to evaluate both personalized and non-personalized recommendation tasks:

\begin{itemize}[leftmargin=5mm]
\item \textbf{MovieLens}. We conduct experiments of personalized item recommendation tasks on the MovieLens-1M dataset~\cite{DBLP:journals/tiis/HarperK16}, a widely used benchmark for evaluating collaborative filtering algorithms. The dataset includes 6,040 users and 3,416 items, with a sparsity of 94.44\%. As in~\cite{DBLP:conf/recsys/HeKM17, DBLP:conf/icdm/KangM18}, we treat all ratings as observed implicit feedback instances, and sort the feedback according to timestamps. For each user, we withhold their last two actions, and put them into validation set and test set respectively. All the rest are used for model training.

\item \textbf{App-to-app Relevance (AAR)}. We also use an app-to-app relevance dataset collected in Google Play Store to evaluate non-personalized item-to-item recommendation~\cite{DBLP:journals/internet/LindenSY03}. The dataset includes over 17M app-to-app relevance scores evaluated by human raters. Each pair is unique, and the relevance score ranges from -100 to 100, indicating how relevant a pair of mobile apps are. This relevance dataset is also highly sparse with a sparsity of 99.98\%. We randomly split the data into 90\% (for training) and 10\% (for evaluation). We seek to build a predictive model to estimate matching scores of unseen app pairs.
\end{itemize}

\subsection{Backbone Recommendation Models}

As quantized embedding is a generic method that can directly replace the embedding layer in existing gradient descent based recommendation models, we include three representative recommendation models as the backbone models to test our hypothesis:
\begin{itemize}[leftmargin=5mm]

\item \textbf{Generalized Matrix Factorization~(GMF)} \cite{DBLP:conf/www/HeLZNHC17}: GMF extends the conventional matrix factorization by introducing a learned linear layer for weighting latent dimensions.

\item \textbf{Neural Matrix Factorization~(NeuMF)} \cite{DBLP:conf/www/HeLZNHC17}:  NeuMF models non-linear interactions between user and item embeddings via multi-layer perceptrons (MLP). NeuMF fuses both the GMF and MLP modules to predict the preference score.

\item \textbf{Self-Attentive Sequential Recommendation~(SASRec)} \cite{DBLP:conf/icdm/KangM18}: The state-of-the-art method on the sequential recommendation task. SASRec adopts multiple self-attention blocks to capture sequential dynamics in users' action history, and predicts the next item at each time step.
\end{itemize}

The above models vary significantly in several ways: 1) depth: GMF is a shallow model while NeuMF and SASRec contains deeper neural networks; 2) architecture: NeuMF adopts MLP while SASRec uses self-attention; 3) target task: GMF and NeuMF are designed for estimating user-item interactions while SASRec seeks to predict the next item based on a sequence of previously viewed items. An illustration of the three models is shown in Figure~\ref{fig:models}.

The embedding dimensionality $d$ is set to 64 for all methods. For NeuMF, we follow the default configurations and adopt a pyramid MLP architecture with 3 hidden layers. We do not use the pre-trained initialization for NeuMF. For SASRec, the maximum sequence length is set to 50, the dropout rate is set to 0.2, and the number of self-attention blocks is set to 2. The number of training epochs is 20 for GMF and NeuMF, and 200 for SASRec, as suggested in the corresponding papers.

\subsection{Recommendation Tasks}
We conduct our experiments on three representative recommendation tasks:
\begin{itemize}[leftmargin=5mm]
\item \textbf{Task 1: Personalized Item Recommendation}. A conventional task seeks to estimate user-item interactions, and thus can be used to generate personalized recommendations for a user (e.g., "Personalized For You" on your homepage). A recommendation model is trained on observed user-item pairs, and seeks to predict items a given user may engage with. In our experiments, we adopt two models: GMF and NeuMF~\cite{DBLP:conf/www/HeLZNHC17}, which are originally designed for this task. We use the MovieLens dataset for this task.

\item \textbf{Task 2: Sequential Recommendation}. Sequential recommendation considers the sequential dynamics in users' action history, and seeks to predict the next item that a user will interact with. To capture sequential patterns, neural sequential models like RNNs and CNNs have been adopted for this task~\cite{DBLP:journals/corr/HidasiKBT15,DBLP:conf/wsdm/TangW18}. We use the self-attention based method SASRec~\cite{DBLP:conf/icdm/KangM18}, which achieves state-of-the-art performance on this task. The experiments are conducted on the MovieLens dataset.

\item \textbf{Task 3: Item to item Recommendation}. A non-personalized recommendation task, which seeks to estimate item-item relevance and is often used for recommending related products (e.g., "Related Products" on the product page). We treat this task as a regression problem, and seek to estimate the app-to-app relevance score in the AAR dataset. We adapt the GMF model into this task via replacing the loss function with a squared loss between estimated scores and ground-truth scores.

\end{itemize}

\begin{table*}

\begin{tabular}{ccccccc}
\toprule
\multirow{2}{*}{Method}                       & \multicolumn{3}{c}{\textbf{GMF}}    & \multicolumn{3}{c}{\textbf{NeuMF}}   \\
                       & HR@10 & NDCG@10 & Model Size & HR@10 & NDCG@10 & Model Size \\
\midrule                      
\textbf{Full Embedding}         & 7.79  & 3.88    & 100\%      & 7.88  & 3.90    & 100\%       \\ \hline
\textbf{Low-rank Factorization (LRF)} &  7.20     &  3.65       &     75.5\%       &   7.41    &      3.70   &   77.3\%         \\
\textbf{Scalar Quantization (SQ)}    &  \underline{7.82}     &  \underline{3.92}       & 25.0\%  & \underline{7.99}  & \underline{3.96}        & 29.0\%           \\

\textbf{DPQ}                    &  \underline{7.92}     &  \underline{3.95}       & 30.4\%    & 7.79       &    \underline{3.90}        &   34.1\%         \\
\textbf{MGQE}                 &  \textbf{\underline{8.03}}    &  \textbf{\underline{3.98}}  &    \textbf{20.7\%}      &  \textbf{\underline{8.04}} &    \textbf{\underline{3.98}} &      \textbf{24.4\%}     \\
\bottomrule
\end{tabular}
\caption{Task 1 - Personalized item recommendation on MovieLens: performance and model size. Underlined numbers indicate reaching the full model performance. All the reported numbers are the average from 10 experiments.}\label{tb:pir}
\vspace{-0.3cm}\end{table*}

\subsection{Compression Approaches}

To test the effectiveness of our embedding compression technique, we compare them with three baselines:
\begin{itemize}[leftmargin=5mm]
\item Baseline 1: \textbf{Full Embedding~(FE)}. The conventional approach which learns a full embedding matrix where each row represents the embedding for an item. This is served as the baseline model.

\item Baseline 2: \textbf{Low-rank Factorization~(LRF)}. A classic approach to reduce parameters in a matrix. We factorize the embedding matrix into two matrices with size $n\times r$ and $r \times d$.

\item Baseline 3: \textbf{Scalar Quantization~(SQ)}. A classic two-step quantization technique\footnote{We use tf.quantization.quantize in TensorFlow API.}. For each dimension in the embedding matrix, SQ records the minimal and maximal values and evenly quantizes the range into $2^{b}$ buckets, where $b$ is the number of bits.

\item \textbf{Differentilable Product Quantization (DPQ)}. DPQ~\cite{DBLP:journals/corr/abs-1908-09756} learns subspace centroids and quantizes the embeddings into the nearest centroid. The quantization procedure is trained end-to-end with the target task.

\item \textbf{Multi-Granular Quantized Embeddings (MGQE)}. The proposed method, designed for power-law distributions. MGQE uses fewer centroids for tail items, which further reduces the model size without a performance drop.

\end{itemize}

For fair comparison, we implement all the methods using \emph{TemsorFlow}. By default, embedding dimension $d$=64 for all methods, the rank $r$ is set to 48 for LRF, the number of bits $b$=8 for SQ, and the number of subspaces $D$=64 and the number of centroids $K$=256 for DPQ. MGQE also uses $D=64$ subspaces, and adopts a two-tier partition where we consider top 10\% items as head items and the rest as tail items. The number of centroids $K_1$ is set to 256 for head items, and $K_2$=64 for tail items. We also show results on more compact configurations for all the methods. All the embeddings are randomly initialized from a normal distribution with a standard deviation of 0.01. All the other hyper-parameters for model training (e.g., learning rate, batch size, etc.) follow the default setting of the backbone recommendation models suggested by the corresponding papers. To reduce the variance, all the reported numbers are the average of the outcomes from 10 experiments.

\subsection{Evaluation Metrics} 

For the personalized recommendation problem, we adopt two common Top-K metrics, Hit Rate@K and NDCG@K, to evaluate recommendation performance~\cite{DBLP:conf/www/HeLZNHC17,DBLP:conf/recsys/HeKM17}. Hit Rate@K counts the fraction of times that the ground-truth next item is among the top K recommended items. NDCG@K is a position-aware metric which assigns larger weights on higher positions. Note that since we only have one test item for each user, Hit Rate@10 is equivalent to Recall@10, and is proportional to Precision@10. For the relevance estimation problem on the AAR dataset, we adopt RMSE (Root Mean Square Error) to evaluate model performance.

We evaluate model size using the number of bits needed to store the model during serving. The model size of full embedding is used as the baseline (i.e., 100\%), and the reported model sizes of compression approaches are normalized correspondingly. Note that we consider the whole model size which includes both the embedding tables and all the other parameters that we do not compress (e.g., weights in hidden layers). However, as the size of embedding tables is typically the dominant term in recommendation models, we can achieve satisfactory compression performance via only compressing the embeddings.

\subsection{Results and Discussions on Compression}

We first choose a relatively large model size for compression techniques, and investigate whether they can achieve the same performance as the baseline (i.e., full embeddings).

\subsubsection{Personalized Item Recommendation} Table~\ref{tb:pir} shows the results of the personalized item recommendation task on the MovieLens dataset. We see that LRF has a significant performance drop, even with modest compression rates. Surprisingly, quantized embeddings (SQ/DPQ/MGQE) generally have performance improvement over the baseline. This may be attributed to the compactness of quantized embeddings, which may alleviate the overfitting problem. We also find that MGQE achieves the best performance with ~20\% of the full model size.

\subsubsection{Sequential Recommendation} Table~\ref{tb:seq} shows the results on compressing the SASRec model. As SASRec leverages the sequential information, it performs much better than GMF and NeuMF. We can see that MGQE maintains the performance with a compression rate of 27.5\%, while other compression approaches generally have a performance drop.

\begin{table}[htp]

\begin{tabularx}{\linewidth}{Cccc}
\toprule
\multirow{2}{*}{\textbf{Method}}         &\multicolumn{3}{c}{\textbf{SASRec}}\\  
                                &\textbf{HR@10}   &\textbf{NDCG@10}   & \textbf{Model Size}  \\\midrule

\textbf{Full Embedding}         &  20.36    & 8.88 & 100\%      \\ \hline
\textbf{LRF}                    &  20.07    & \textbf{\underline{9.21}} & 74.3\%      \\
\textbf{Scalar Quantization}    &  20.04    & 8.72 & 31.5\%      \\
\textbf{DPQ}                    &  20.23    & 8.81 &38.3\%\\
\textbf{MGQE}                  &  \textbf{\underline{20.36}}    & \underline{8.88} & \textbf{27.5\%}      \\
\bottomrule
\end{tabularx}
\caption{Task 2 - Sequential Recommendation on Movielens: performance and model size.}\label{tb:seq}
\vspace{-0.3cm}\end{table}

\subsubsection{Item-to-item Recommendation} The results of RMSE on the AAR dataset are shown in Table~\ref{tab:aar}. We can see that SQ has a performance drop, while DPQ and MGQE significantly improve the performance. Presumably quantized embeddings can regularize the model and alleviate overfitting, and thus we observe performance improvement on the highly sparse dataset AAR. This also shows that MGQE is able to maintain (or improve) the full model performance not only on ranking problems (which considers relative values) but also on regression problems (which needs absolute estimations).

\begin{table}[htp]

\begin{tabularx}{\linewidth}{Ccc}
\toprule
\multirow{2}{*}{\textbf{Method}}         &\multicolumn{2}{c}{\textbf{AAR}}\\  
                                &\textbf{RMSE}       & \textbf{Model Size}  \\\midrule

\textbf{Full Embedding}         & 30.34     & 100\%      \\ \hline
\textbf{Scalar Quantization}    & 30.42     & 25.0\%      \\
\textbf{DPQ}                    & \underline{29.57}     & 25.1\%      \\
\textbf{MGQE}                  & \textbf{\underline{29.49}}     & \textbf{19.4\%}      \\
\bottomrule
\end{tabularx}
\caption{Task 3 - Item-to-item Recommendation on AAR data: performance and model size. Lower is better for RMSE.}
\label{tab:aar}
\vspace{-0.3cm}\end{table}

\begin{figure*}
\centering
\begin{subfigure}[b]{0.38\textwidth}
\includegraphics[width=\linewidth]{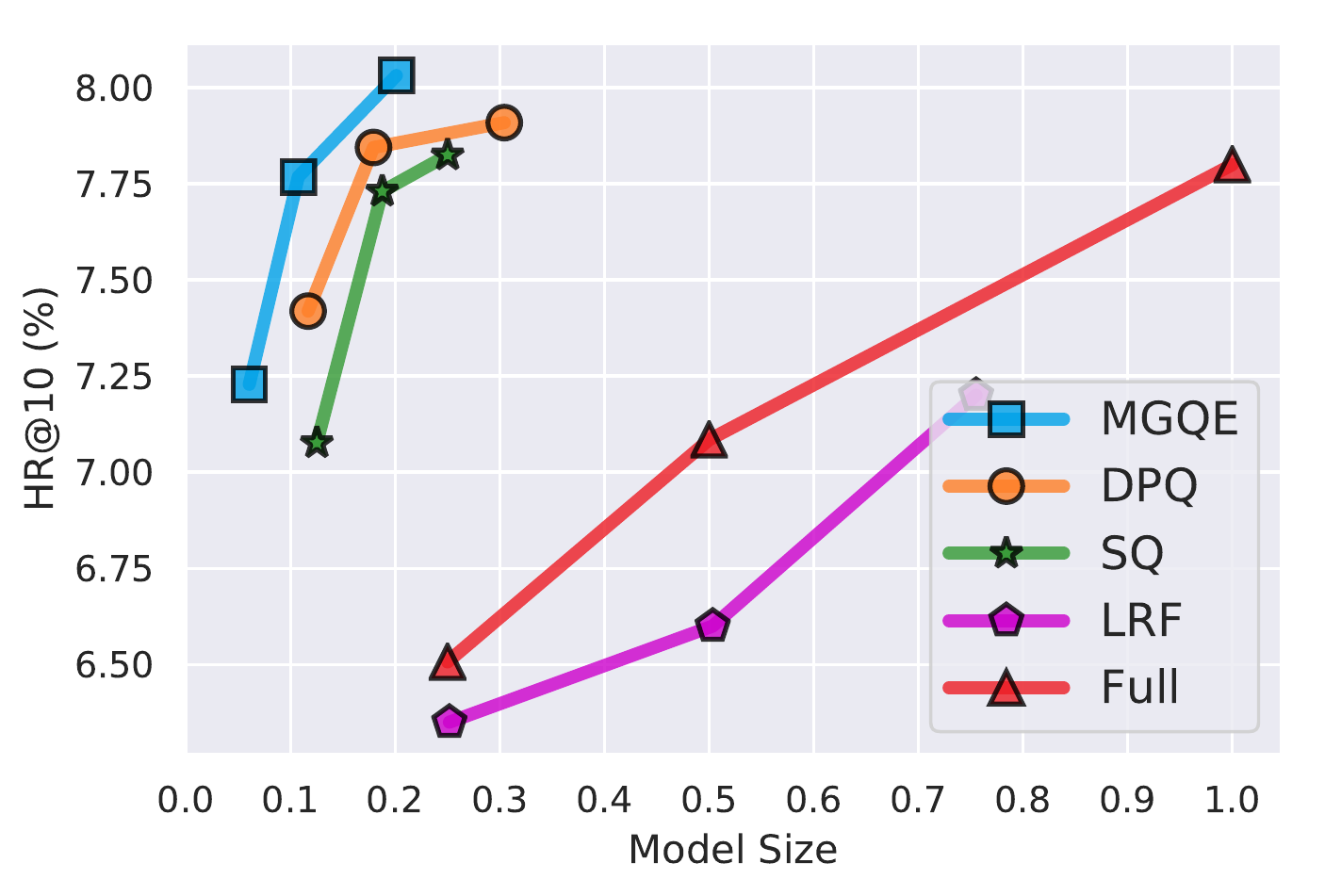}
\subcaption{Task 1: MovieLens (GMF)}
\end{subfigure}
\hspace{0.08\textwidth}
\begin{subfigure}[b]{0.38\textwidth}
\includegraphics[width=\linewidth]{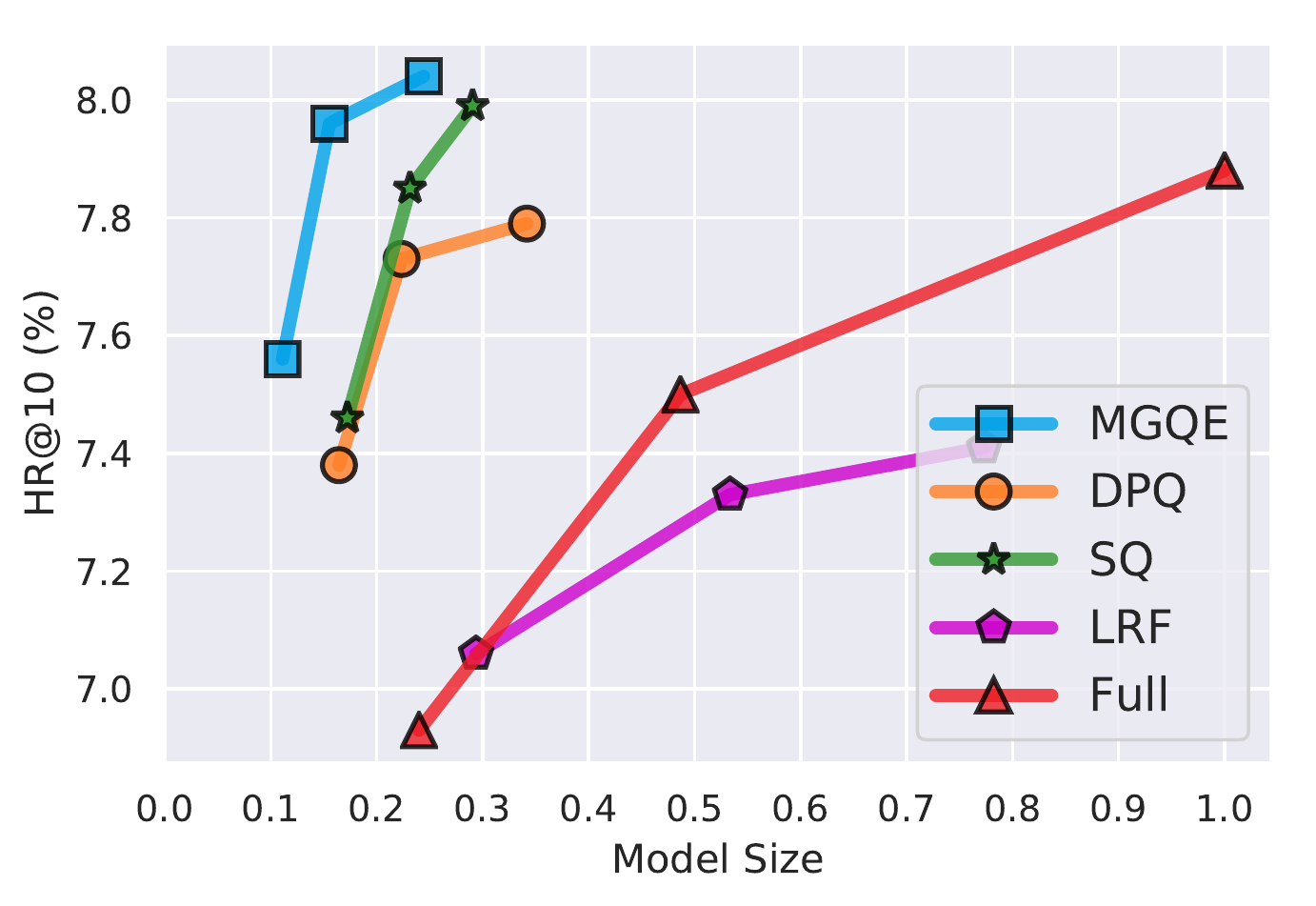}
\subcaption{Task 1: MovieLens (NeuMF)}
\end{subfigure}
\begin{subfigure}[b]{0.38\textwidth}
\includegraphics[width=\linewidth]{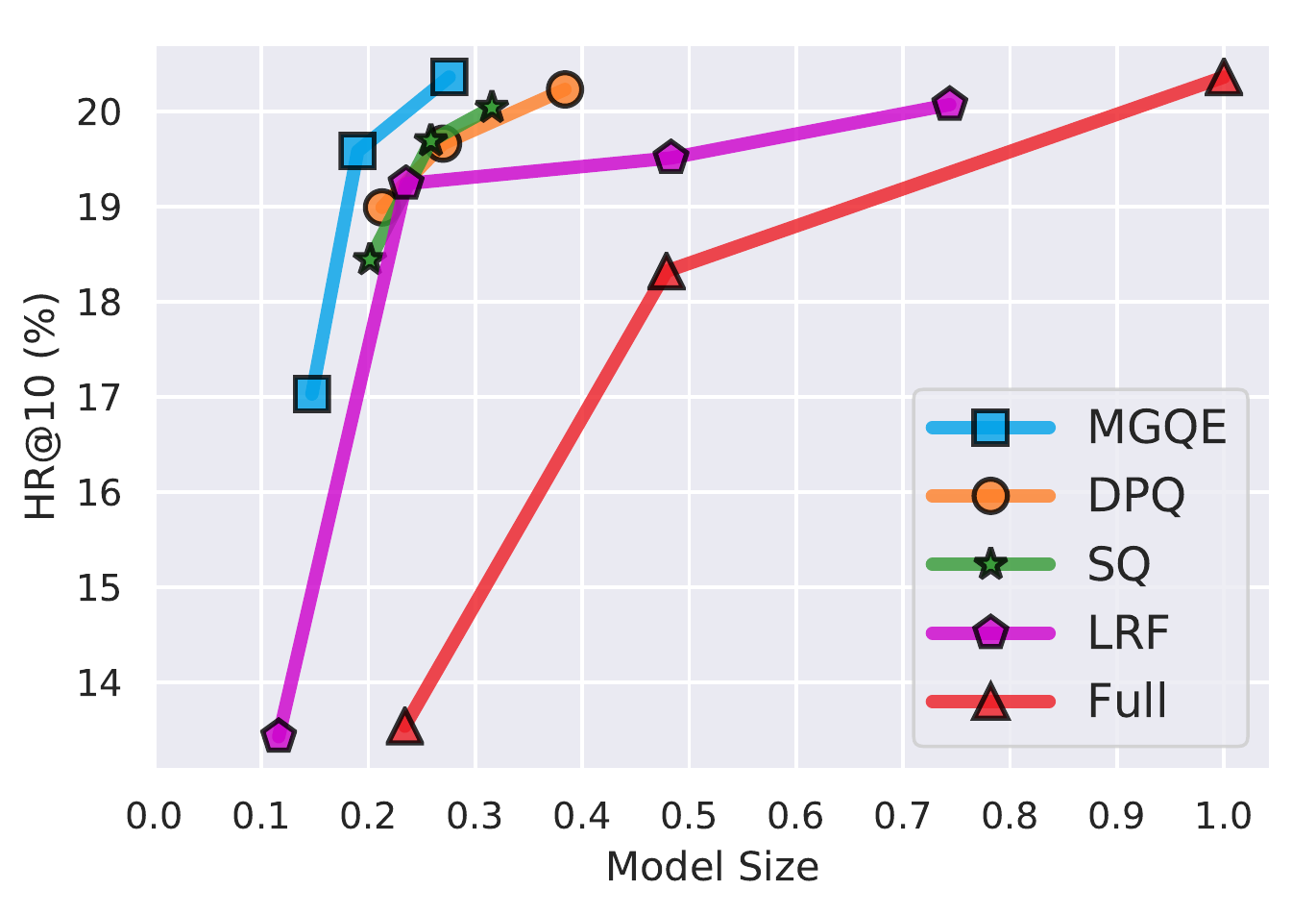}
\subcaption{Task 2: MovieLens (SASRec)}
\end{subfigure}
\hspace{0.08\textwidth}
\begin{subfigure}[b]{0.38\textwidth}
\includegraphics[width=\linewidth]{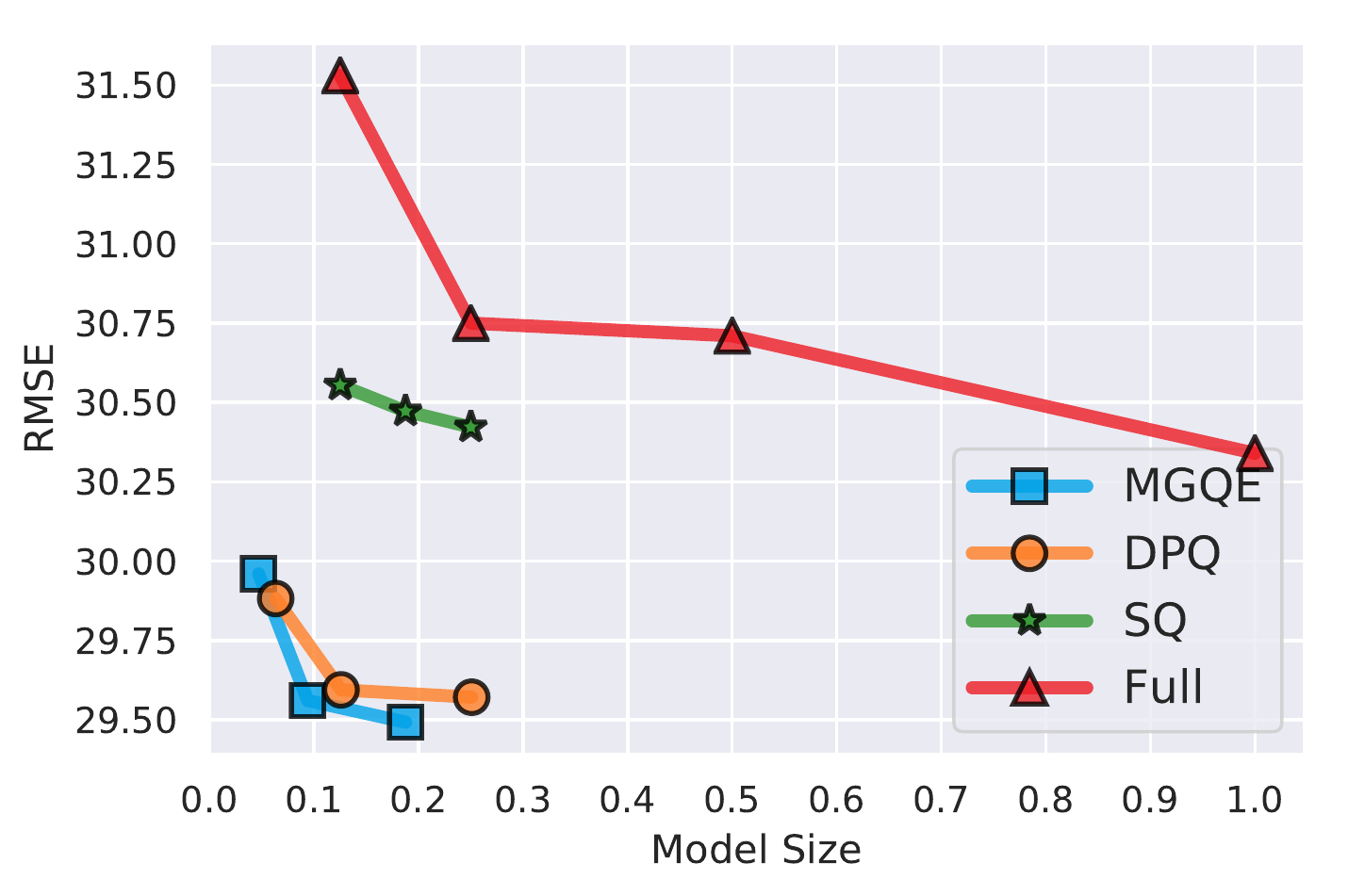}
\subcaption{Task 3: AAR (GMF)}
\end{subfigure}

\caption{Performance of embedding compression methods with different model sizes.}\label{fig:model_size}
\end{figure*} 
\subsubsection{The Effect of Model Size} Figure~\ref{fig:model_size} shows the performance with more compact model sizes on the four tasks. For full embeddings, we vary the dimensionality $d$ to adjust its model sizes. For scalar quantization, we vary the number of bits per dimension. For DPQ/MGQE, we vary the number of subspaces $D$. We can see that the performance of full embeddings drops significantly with smaller model size, which shows directly reducing the dimensionality is not an effective way to compress recommendation models. With the same model size, MGQE has the best performance; and with the same recommendation performance, MGQE has the smallest model size. Hence, MGQE is an effective embedding learning and compression method, and can be applied to achieve better performance and compress recommendation models.


\subsubsection{Summary}
Based on the results of applying embedding compressing methods for three recommendation models~(GMF, NeuMF, SASRec) on two datasets~(MovieLens and AAR), we found (1) DPQ matches the performance of full embeddings in most cases; (2) MGQE matches (and sometimes improves) the performance with full embeddings in all cases. This verifies the \textbf{RQ1} that quantized embeddings are able to reduce the model size while matching or even improving the model performance. Moreover, we found MGQE generally outperforms baselines under different compression ratios. This verifies the \textbf{RQ2} that MGQE outperforms alternative compression approaches in recsys tasks.

\begin{figure*}
    
    \centering
    \begin{subfigure}[b]{.33\linewidth}
    \includegraphics[width=\linewidth]{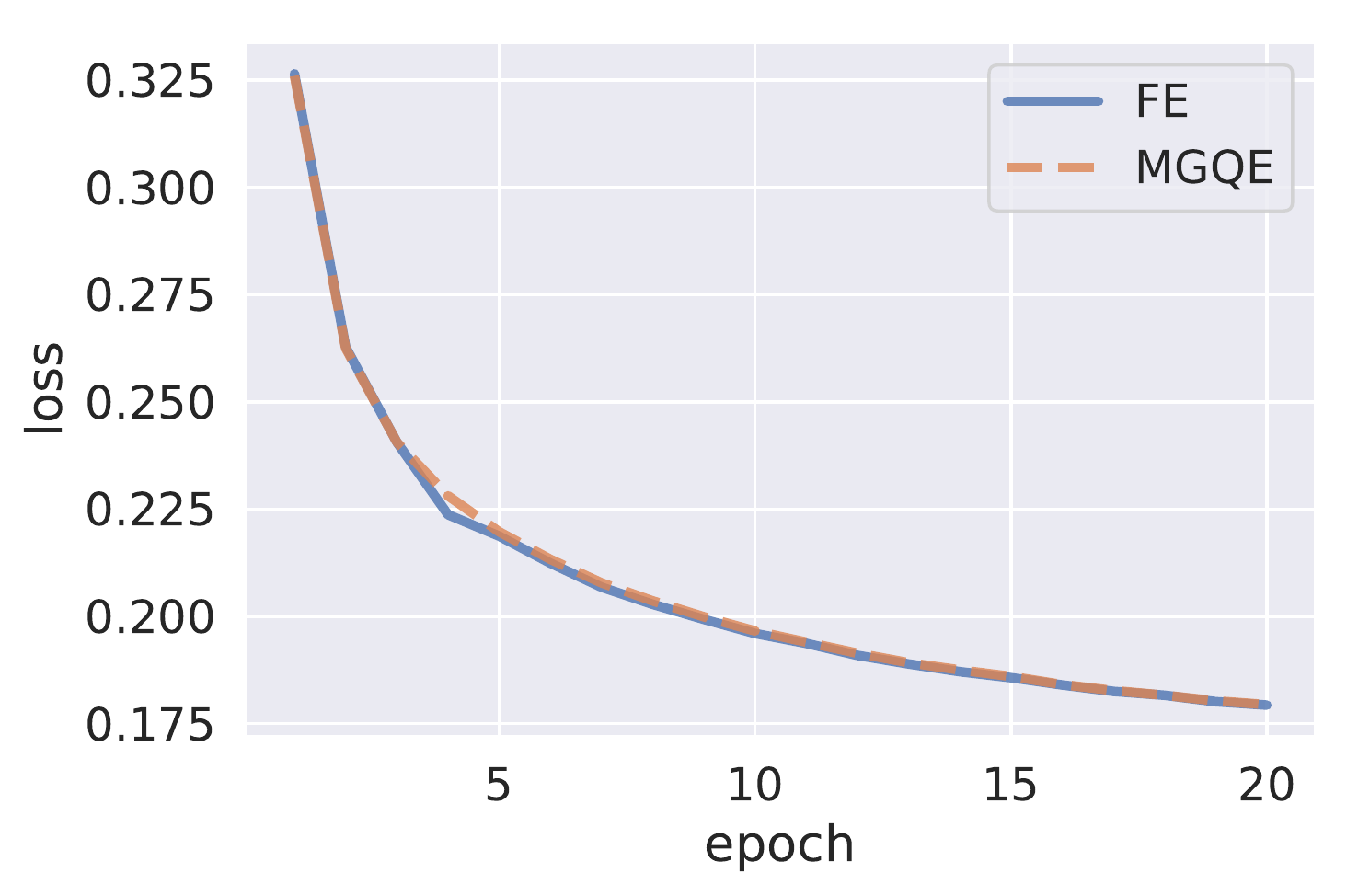}
    \subcaption{Task 1: GMF}
    \end{subfigure}
    \centering
    \begin{subfigure}[b]{.33\linewidth}
    \includegraphics[width=\linewidth]{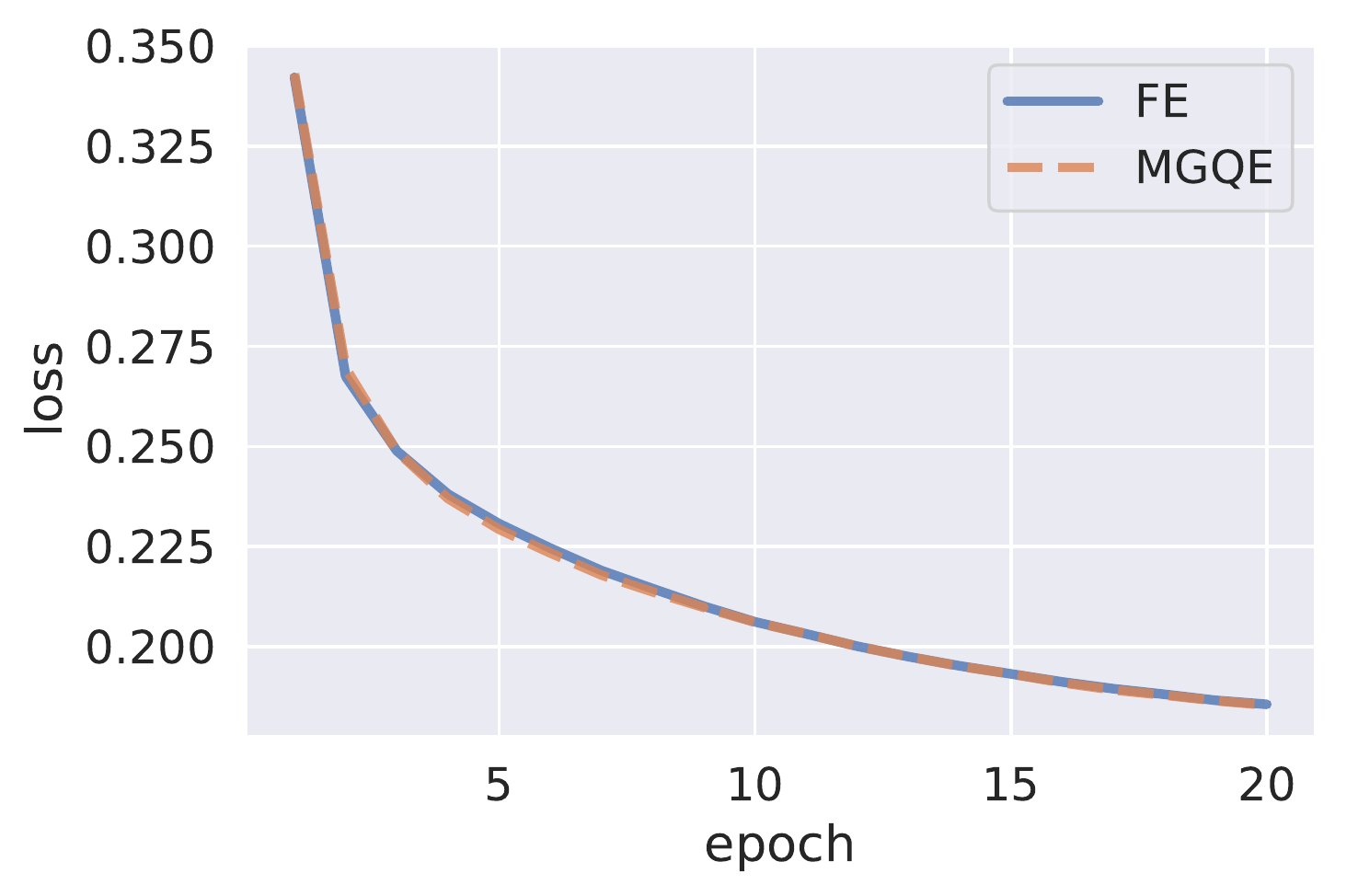}
    \subcaption{Task 1: NeuMF}
    \end{subfigure}
    \centering
    \begin{subfigure}[b]{.33\linewidth}
    \includegraphics[width=\linewidth]{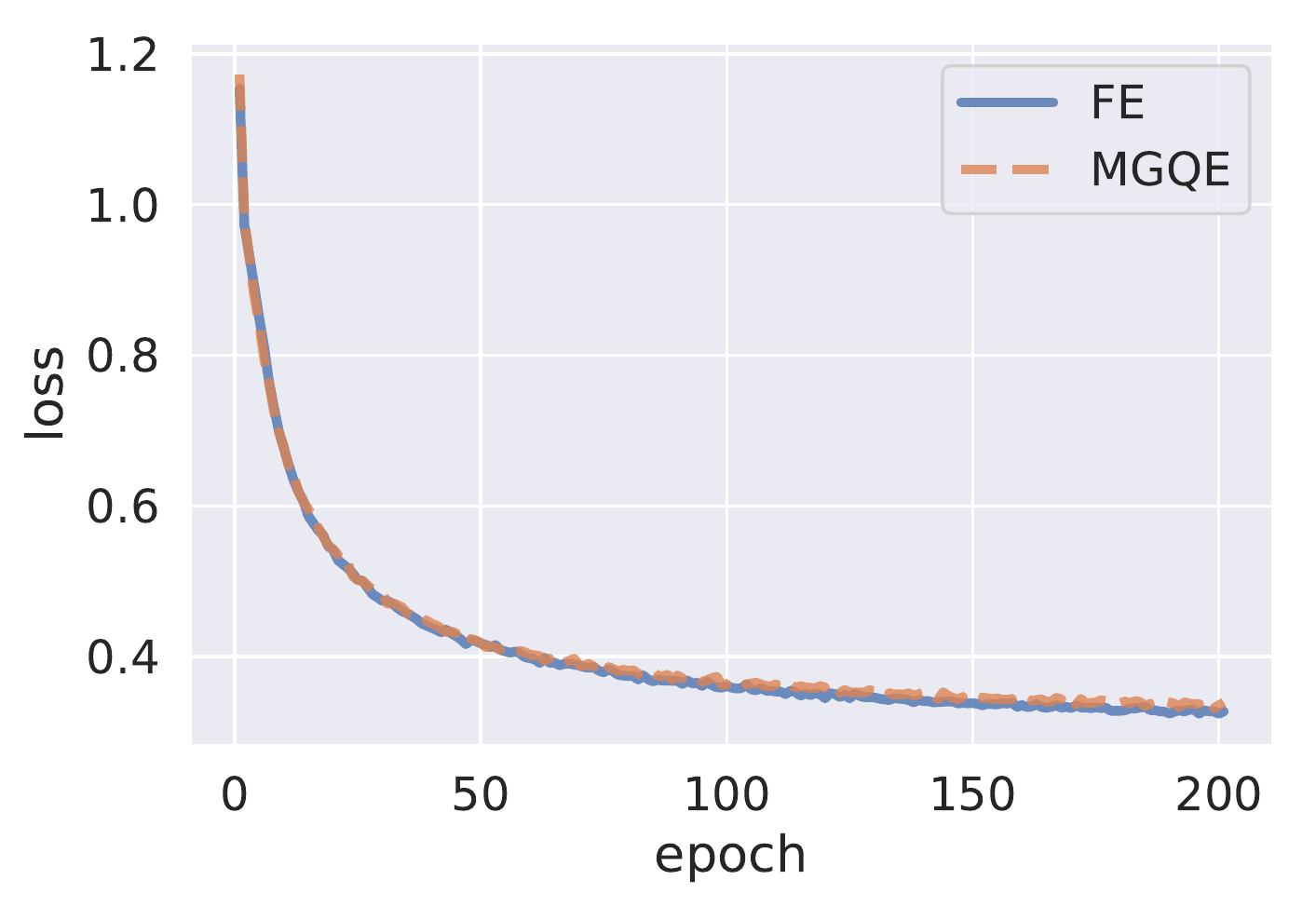}
    \subcaption{Task 2: SASRec}
    \end{subfigure}
    \caption{Training loss of MGQE and full embeddings~(FE) on the three recommendations models on Movielens.}
    \label{fig:curves}
\end{figure*}
\subsection{The Choices of Multi-granularity}

We investigate \textbf{RQ3} by evaluating the three MQGE variants on the MovieLens dataset, and the results are shown in Table~\ref{tab:variant}. We see that the default approach with shared centroids achieves the same (or better) performance as (than) the full model performance on all models. Hence we choose this variant as the default approach. The variant with unshared centroids and varying numbers of centroids $\widetilde K$ is the runner up variant, and outperforms the default variant with NeuMF. This might be attributed to its larger model size, as it maintains two centroid embedding tables. However, using unshared centroids costs additional storage space, hence we choose the shared variant as the default for MGQE.

\begin{table}[htp]
\begin{tabularx}{\linewidth}{Ccc}
\toprule
\textbf{Variant}  &\textbf{HR@10}   &\textbf{NDCG@10}   \\\midrule
\multicolumn{3}{c}{\textbf{GMF}}\\
\textbf{Full Embedding}         &   7.79     &   3.88  \\        
\textbf{unshare, $D=64$, $\widetilde K=[256, 64]$}       &  \underline{7.79}     &   3.85           \\
\textbf{unshare, $\widetilde D=[64, 32]$, $K=256$}        &  \underline{7.89}     &   \underline{3.96}           \\
\textbf{share, $D=64$, $\widetilde K=[256, 64]$}         &  \underline{\textbf{8.03}}     &   \underline{\textbf{3.98}}           \\
\hline
\multicolumn{3}{c}{\textbf{NeuMF}}\\
\textbf{Full Embedding}         &   7.88     &   3.90  \\
\textbf{unshare, $D=64$, $\widetilde K=[256, 64]$}       &  \underline{\textbf{8.22}}     &   \underline{\textbf{4.04}}           \\
\textbf{unshare, $\widetilde D=[64, 32]$, $K=256$}        &  7.66     &   3.72           \\
\textbf{share, $D=64$, $\widetilde K=[256, 64]$}         &  \underline{8.04}     &   \underline{3.98}           \\
\hline
\multicolumn{3}{c}{\textbf{SASRec}}\\
\textbf{Full Embedding}         &   20.36     &   8.88  \\
\textbf{unshare, $D=64$, $\widetilde K=[256, 64]$}       &  20.14     &   8.84           \\
\textbf{unshare, $\widetilde D=[64, 32]$, $K=256$}        &  20.27     &   8.83           \\
\textbf{share, $D=64$, $\widetilde K=[256, 64]$}         &  \underline{\textbf{20.36}}     &   \underline{\textbf{8.88}}           \\
\bottomrule
\end{tabularx}
\caption{Performance of MGQE variants on the MovieLens dataset. Underlined numbers indicate reaching the full model performance.}\label{tab:variant}
\vspace{-0.3cm}\end{table}

\begin{figure}[t]
    \centering
    \begin{subfigure}[b]{.75\linewidth}
    \includegraphics[width=\linewidth]{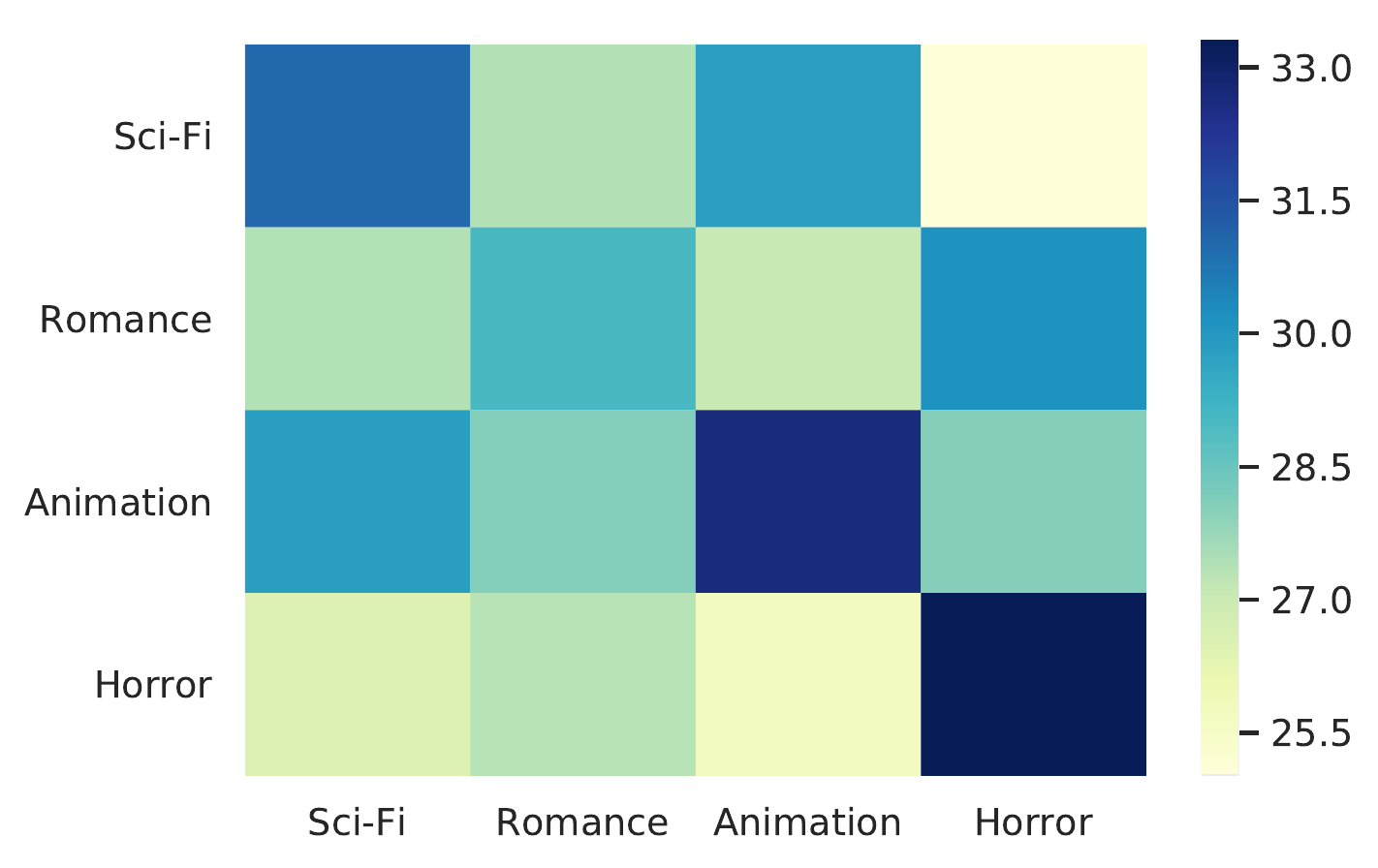}
    \subcaption{SQ}
    \end{subfigure}
    \begin{subfigure}[b]{.75\linewidth}
    \includegraphics[width=\linewidth]{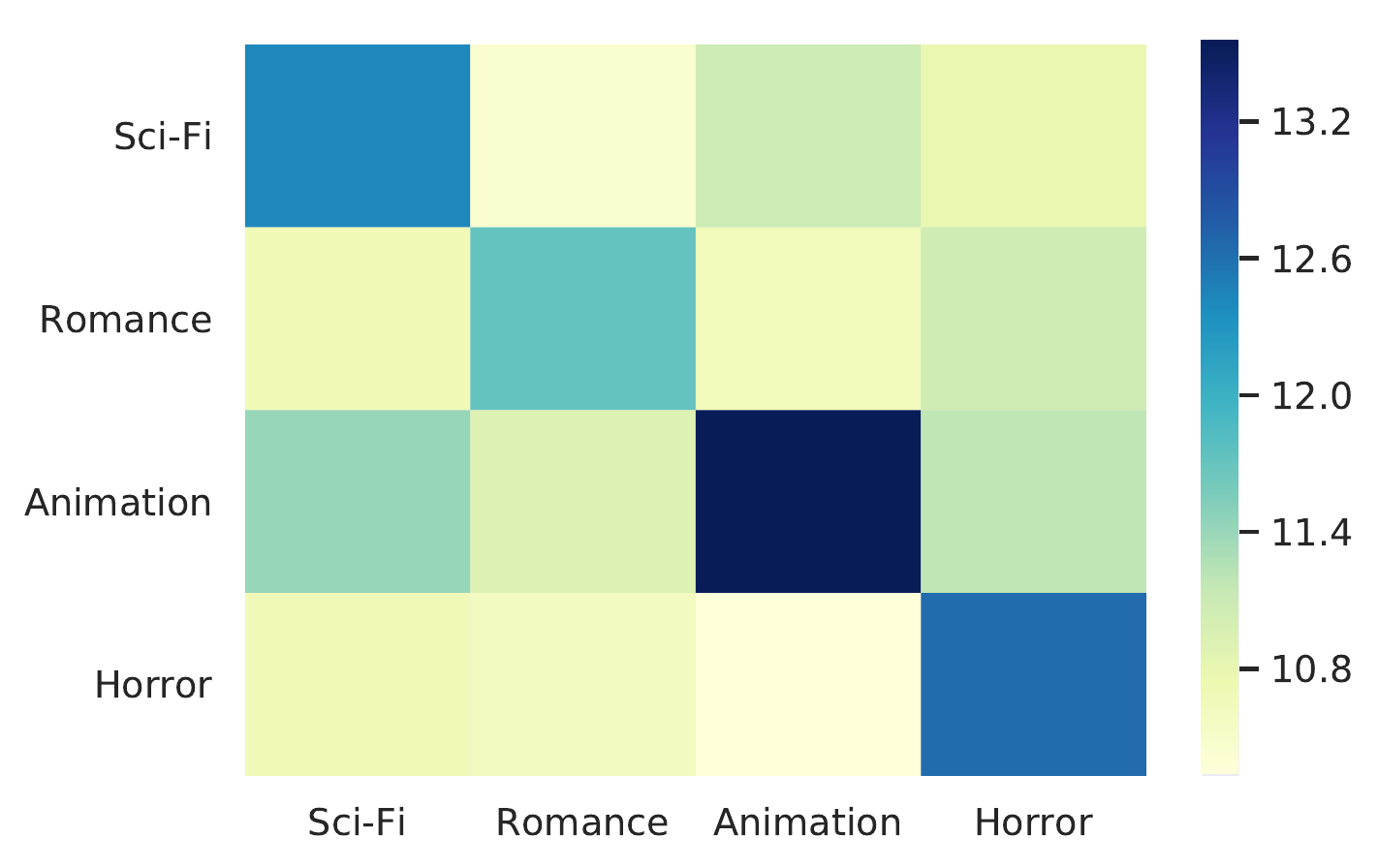}
    \subcaption{MGQE}
    \end{subfigure}
    \caption{Visualization of the average code similarity between movies from the four categories. Darker color means higher code similarity.}
    \label{fig:vis}
\end{figure}

\subsection{Convergence}

As MGQE can directly replace the full embedding layers, it is important to investigate the training of MGQE to check whether the optimization process is the same as or similar to the original one. By doing this, we can check some potential issues like unstable training or slow convergence speed due to the discrete and compact embedding structure. Figure~\ref{fig:curves} shows the training curves of MGQE on the three recommendation models on the MovieLens dataset, compared with full embeddings. As in our other experiments, we use the default hyper-parameters (e.g., learning rate, batch size, etc.) that are originally designed for full embeddings. We can see that the training processes of MGQE are stable, and closely approximate that of full embeddings. This verifies \textbf{RQ4} that MGQE behaves similarly to full embeddings in terms of convergence trajectories.

\subsection{Visualization}

To investigate \textbf{RQ5} (whether the learned quantization is able to cluster similar items), we examine the codes from scalar quantization and MGQE on GMF. In the MovieLens dataset, each movie has several categories (though not used for training), we randomly select two disjoint sets where each set contains 200 movies from 4 categories: Science Fiction (Sci-Fi), Romance, Animation, and Horror. Intuitively, movies from the same category should have similar codes. We evaluate the code similarity\footnote{We count the number of positions that two codes share the same centroid. Hence the similarity ranges from 0 to $D$.} of any two movies and show the average similarity between categories in Figure~\ref{fig:vis}. We can see that both SQ and MGQE tend to use similar codes to represent similar movies, as the diagonal values are relatively large. However, for dissimilar movies, SQ can not distinguish them very well, while MGQE assigns significantly low similarities between dissimilar movies. This shows that, compared with the two-step quantization method SQ, the end-to-end quantization helps MGQE learn better centroids for the target task.



\section{Conclusions and future work}

We investigated the embedding compression problem for recommender systems, and proposed multi-granular quantized embeddings~(MGQE) for compressing large-vocabulary categorical features. MGQE adopts differentiable quantized representations to reduce the model size, and further cuts down the storage space by using fewer centroids for tail items. MGQE is a generic approach that can be used to replace the embeddings layers in existing recommendation models, and trained end-to-end for the target task. We conducted extensive experiments on compressing three representative recommendation models for three different recommendation tasks. Our results show that MGQE outperforms the baselines, and reaches the full model performance with nearly 20\% of the full model size. 
In the future, we plan to investigate (i) learned fine-grained item partitions for multi-granular capacity allocation; (ii) jointly quantized embeddings for multiple categorical features; and (iii) quantized neural network weights for recommendation models.

\bibliographystyle{ACM-Reference-Format}
\bibliography{MGQE}


\end{document}